\RequirePackage{fix-cm}
\documentclass[twocolumn]{svjour3}          
\smartqed  
\usepackage{graphicx}
\usepackage{bmpsize}
\graphicspath{ {Figures/} }

\usepackage{supertabular} 
\usepackage{natbib}
\usepackage{subcaption}
\usepackage{amsmath}
\journalname{Experiments in Fluids}
\begin{document}

\title{Insights into Leading Edge Vortex Formation and Detachment on a Pitching and Plunging Flat Plate}


\author{Johannes Kissing$^1$$^{\dagger}$ \and Jochen Kriegseis$^2$ \and Zhenyao Li$^3$ \and Lihao Feng$^3$ \and Jeanette Hussong$^1$ \and Cameron Tropea$^1$ \thanks{\noindent$^{\dagger}$Corresponding author. \email{kissing@sla.tu-darmstadt.de}}}


\authorrunning{J. Kissing, J. Kriegseis, Z. Li, L. Feng, J. Hussong \& C. Tropea} 

\institute{$^1$Institute of Fluid Mechanics and Aerodynamics\\
Technische Universit\"at Darmstadt\\
Flughafenstrasse 19\\
64347 Griesheim, Germany\\
$^2$Institute of Fluid Mechanics\\
Karlsruhe Institue of Technology\\
Kaiserstr. 12\\
76131 Karlsruhe, Germany\\
$^3$Department of Fluid Mechanics\\
Beihang University\\
No. 37 Xueyuan Road, Haidian District\\
Beijing, P.R. China, 100083
}

\date{Received: date / Accepted: date}

\maketitle

\begin{abstract}

\qquad The present study is a prelude to applying different flow control devices on pitching and plunging airfoils with the intention of controlling the growth of the leading edge vortex (LEV); hence, the lift under unsteady stall conditions. As a pre-requisite, the parameters influencing the development of the LEV topology must be fully understood, and this constitutes the main motivation of the present experimental investigation. The aims of this study are twofold.

First, an approach is introduced to validate the comparability between flow fields and LEV characteristics of two different facilities using water and air as working media by making use of a common baseline case. The motivation behind this comparison is that with two facilities the overall parameter range can be greatly expanded. This comparison includes an overview of the respective parameter ranges, control of the airfoil kinematics and careful scrutiny of how post-processing procedures of velocity data from time-resolved particle image velocimetry (PIV) influence the integral properties and topological features used to characterise the LEV development. 

Second, and based on results coming from  both facilities, the appearance of secondary structures and their effect on LEV detachment over an extended parameter range is studied. A Lagrangian flow field analysis, based on finite-time Lyapunov Exponent (FTLE) ridges, allows  precise identification of secondary structures and reveals that their emergence is closely correlated to a vortex Reynolds number threshold computed from the LEV circulation. This threshold is used to model the temporal onset of secondary structures. Further analysis indicates that the emergence of secondary structures causes the LEV to stop accumulating circulation if the shear layer angle at the leading edge of the flat plate has ceased to increase. This information is of particular importance for  advanced control circuit application, since  efforts to strengthen and/or prolong LEV growth  rely on precise knowledge about where and when to apply flow control measures.

\end{abstract}


\section{Introduction}
\label{sec:sec_Intro}
High lift at low Reynolds numbers is an essential feature of biological propulsion based on flapping wings and is a promising technology for future micro air vehicle (MAV). In terms of dimensionless parameters, hovering insects and birds in cruise flight were found to execute wing kinematics at a chord based Reynolds number $Re=U_{\infty} c / \nu $ of the order of $10^{3}-10^{4}$, where $U_\infty$ is the free-stream velocity, $c$ the airfoil chord and $\nu$ is the kinematic viscosity \citep[cf.][]{Ellington.1984}. The Strouhal number $St = 2 f h / U_{\infty}$, where $f$ is the plunging frequency and $h$ the plunging amplitude, varies between 0.2 and 0.4 for efficient propulsion, while the reduced frequency $k= \pi c f / U_{\infty}$ is optimized according to the respective $St$ and the wing planform area \cite[cf.][etc.]{Triantafyllou.1993, Nudds.2004} MAVs are designed for a wide range of $Re$ from 0 up to 60,000, a higher $k$ between 0.146 and 1.2 and lower $St$ up to 0.07, compared to biological flapping flight, due to the high inertial loads of moving wings \cite[cf.][]{Jones.2010, Jones.2000, Croon.2015}.

High transient lift on flapping wings of insects and birds is attributed to  leading edge vortex (LEV) growth on the wing \cite[e.g.][]{Ellington.1996}. It occurs when the effective angle of attack of the inflow on the airfoil ($\alpha_{\mathrm{eff}}$) changes dynamically, such that the leading edge shear layer separates due to an adverse pressure gradient and subsequently rolls up into a vortex. This process, known as dynamic stall \cite[cf.][]{CARR.1988}, leads to a collapse of the induced lift as soon as the LEV detaches from the airfoil and is convected downstream. Therefore, maintaining a longer vortex growth phase by delaying its detachment with the aid of local flow control can increase the overall vortex induced lift on flapping wings and thus enhance the manoeuvrability as well as the gust-tolerance of MAVs \cite[cf.][]{Eldredge.2019}.

The long-term objective of the current study is  manipulation of the flow field around the LEV on a pitching and/or plunging flat plate in order to attain higher overall transient lift by delaying the LEV detachment or increasing its circulation. The idea is to manipulate the flow field at topologically critical locations using a dielectric barrier discharge plasma actuator (DBD-PA) in air and a synthetic jet actuator (SJ) in water. To enable this manipulation  the underlying mechanisms have to first be sufficiently understood. 

\begin{figure}[h]
    	\includegraphics[width=\columnwidth, trim = 0 0 0 0, clip]{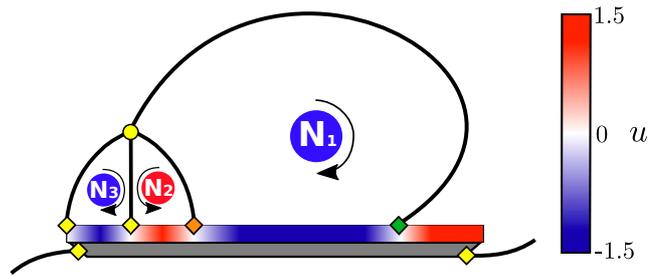}
        \caption{Sketch of the flow topology during leading edge vortex growth on an unsteady flat plate. Half-saddles are marked with diamonds and full saddle with a circle. The main LEV (node N\textsubscript{1}) and secondary vortices (nodes N\textsubscript{2} and N\textsubscript{3}) are highlighted in addition to the tangential velocity on the airfoil surface $u$ induced by them. Adapted from \cite{Rival.2014}.}
        \label{fig:Topology}
\end{figure}

\cite{Rival.2014} found the chord length $c$ to be the characteristic length scale for vortex detachment on a plunging flat plate with different leading edge geometries and a NACA 0012 airfoil. By considering the flow topology, following concepts introduced by \cite{Foss.2004}, they found that the LEV induced lift on the airfoil drops when fluid begins to recirculate around the trailing edge. An early stage of the flow topology during the LEV growth phase on an unsteady flat plate is depicted in Fig.\ \ref{fig:Topology}.
Here, the LEV is denoted as node N\textsubscript{1}. Recirculation is initiated when the rear stagnation point behind the LEV on the airfoil surface, marked by a green diamond in Fig. \ \ref{fig:Topology}, travels beyond the trailing edge. The recirculated fluid is entrained between the vortex and airfoil, and finally feeds secondary vortices (nodes N\textsubscript{2} and N\textsubscript{3}) ahead of the main vortex, which grow and interrupt the LEV from its feeding shear layer.

Nevertheless, \cite{Sattari.2012} found in a generic experiment that a two-dimensional starting vortex produced by a DBD-PA on the rear edge of a plate detaches from its feeding shear layer in absence of any geometric length scale. This suggests that vortex detachment can occur independent of any length scale. Likewise, \cite{Widmann.2015} found that for a pitching and plunging flat plate at intermediate $Re$ and a higher $k$, in combination with a large effective angle of attack amplitude $\hat{\alpha}_{\mathrm{eff}}$, the LEV stops accumulating circulation before the rear stagnation point behind the vortex reaches the trailing edge. They conclude that $c$ is not the defining length scale for the investigated experimental parameters. Instead a viscous response of the boundary layer between the LEV and the airfoil is identified to cause an abrupt eruption of surface fluid that initiates the growth of secondary vortices ahead of the main vortex. In this detachment scenario these secondary vortices, also referred to as secondary structures, grow and cut off the LEV from its feeding shear layer. From a topological point of view, growing secondary structures (nodes N\textsubscript{2} and N\textsubscript{3} in Fig.\ \ref{fig:Topology}) cause their rear confining stagnation point, marked as an orange diamond in in Fig.\ \ref{fig:Topology}, to travel downstream and finally merge with the LEV confining stagnation point (green diamond). They term this locally initiated detachment mechanism 'boundary-layer eruption', which adopts the terminology used by \cite{Doligalski.1994}. This kind of detachment without recirculation of fluid around the trailing edge, where $c$ is not the characteristic length scale, was also observed by \cite{EslamPanah.2015} and \cite {Akkala.2017} for an LEV on a plunging flat plate at high $k$ between 1 and 2.

A major objective of the present study is to establish a basis for future flow control attempts on LEVs by using two different flow actuation mechanisms working with different media. Therefore, experiments were carried out  both in water and air under geometric, kinematic and dynamic similarity. The choice of different working media is not only related to flow manipulation devices used, but also allows an extended dimensionless parameter range. At first, a common baseline case is defined  to enable comparability of the flow field and vortex characteristics between results from the two facilities. Second, with the intention to better understand the formation of secondary structures during LEV growth as well as their consequences regarding the LEV detachment process, the topology of the flow field for different dimensionless parameters and effective angle of attack amplitudes is investigated. Additionally, the emergence of secondary structures is modelled to allow for a precise timing of flow control approaches targeting secondary structure manipulation.

\section{Facilities and Methods}
\label{sec:sec_FacMeth}
\subsection{Parameter Space and Facilities}
\label{SubSec:FacPar}
Experimental investigations of the LEV formation and detachment on a pitching and plunging flat plate were conducted at two different facilities: a wind tunnel at the Technische Universit\"at Darmstadt (TUDA) and a water tunnel at Beihang University (BUAA). Both set-ups use time-resolved particle image velocimetry (PIV) to characterize the flow field.

At TUDA an open return wind tunnel with a test section of 0.45 m $\times$ 0.45 m was used, whereby the turbulence level measured by hot-wire anemometry was found to be less than 2\,\% for $U_{\infty} = 3.45 \pm 0.05$\,m/s. At BUAA a water tunnel with a test section of 1 m $\times$ 1.2\,m in spanwise and vertical orientation was used in which the RMS turbulence level was found to be less than 1.3\,\% of the free-stream velocity investigated in this study. The investigated airfoil at both facilities was a flat plate of less than 6\,\% thickness and $c=120$\,mm with a sharp leading edge of $30 \, ^{\circ}$ in order to produce a defined separation of the leading edge shear layer.

\begin{figure}[h]
    	\includegraphics[width=\columnwidth, trim = 5 5 35 3, clip]{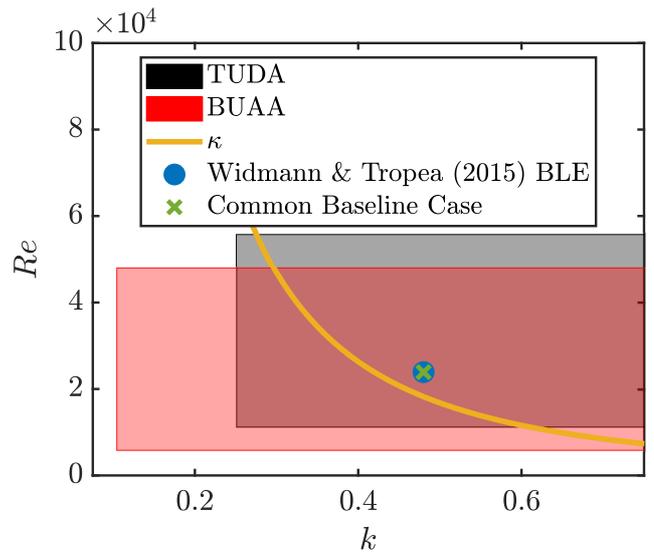}
        \caption{Dimensionless parameter ranges of facilities at BUAA and TUDA for $St = 0.1$, $\hat{\alpha}_{\mathrm{eff}} = 30 \, ^{\circ}$ and $c = 120$\,mm. The covering ratio $\kappa$ introduced by \cite{Widmann.2015}, differentiating between detachment mechanisms, is indicated with a curve in addition to their condition where boundary-layer eruption detachment (BLE) was observed. Finally, the chosen common baseline case for this study is shown.}
        \label{fig:ParameterRange}
    \end{figure}

Highly dynamic pitch and plunge kinematics of the flat plate airfoil require \textit{a priori} consideration of experimental capabilities in terms of attainable dimensionless parameters at both facilities to identify a common parameter space. Limiting factors for the maximum reduced frequency and Strouhal number are the maximum available actuator force to move the airfoil and maximum allowable plunge height, considering the airfoil mass and chord length. The $Re$ range is determined by the lowest and highest free-stream velocity attainable in the respective tunnels, in relation to the chord length, while $U_{\infty}$ also influences the $k$ and $St$ ranges. Fig.\ \ref{fig:ParameterRange} illustrates the attainable parameter ranges in the respective facilities and their overlap for $St=$ 0.1 and $\hat{\alpha}_{\mathrm{eff}} = 30 \, ^{\circ}$ with $c = 120$\,mm.
The dimensionless and geometric parameters of the common baseline case have been chosen such that a boundary-layer eruption detachment mechanism of the LEV can be expected, as discussed in section~\ref{sec:sec_Intro}. \cite{Widmann.2015} developed an analytical parameter, which enables the identification of dimensionless parameters that lead to  boundary-layer eruption detachment, termed the dimensionless covering ratio $\kappa$ as indicated in Fig.\ \ref{fig:ParameterRange}. For cases with parameters located above this line, the LEV should detach due to boundary-layer eruption. One case where a boundary-layer eruption was identified at $k=$ 0.48, $Re=$ 24,000, $St=$ 0.1 and $\hat{\alpha}_{\mathrm{eff}} =$ 30$\, ^{\circ}$ for $c=$ 120\,mm is depicted with a blue dot. This is chosen as the common baseline case for both facilities in the current study.

\begin{figure}[h]
    	\includegraphics[width=\columnwidth, trim = 0 0 0 0, clip]{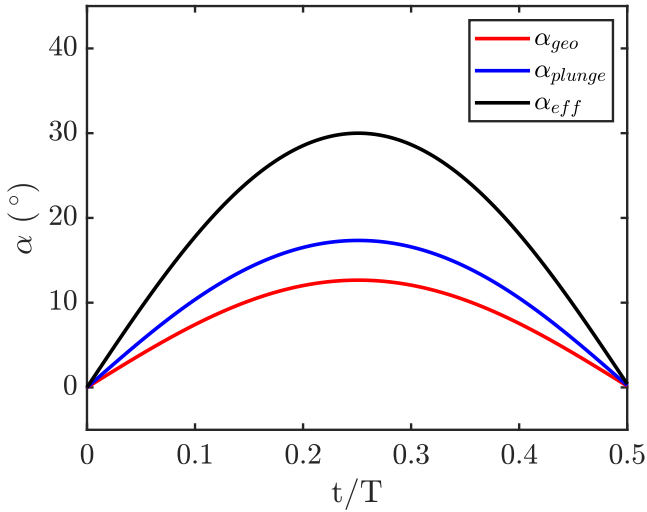}
        \caption{Evolution of the effective angle of attack $\alpha_{\mathrm{eff}}(t)$ during the downstroke as a result of the superposition of the induced angle of attack due to the plunging motion $\alpha_{\mathrm{plunge}}(t)$ and the  geometric angle of attack $\alpha_{\mathrm{geo}}(t)$ as a function of  dimensionless time ($t/T$)  for the common baseline case.}
        \label{fig:AlphaEff}
    \end{figure}

Apart from dimensionless parameters, the effective inflow angle evolution on the airfoil $\alpha_{\mathrm{eff}}(t)$ is an important parameter determining the LEV development, since it affects the shear layer and LEV characteristics through the vertical inflow velocity component on the airfoil. The magnitude of $\alpha_{\mathrm{eff}}(t)$, $\hat{\alpha}_{\mathrm{eff}}$, is determined by the  addition of the inflow angle induced by the plunging motion $\alpha_{\mathrm{plunge}}(t) = \dot{h}(t) / U_{\infty}$ and the  geometric angle of the airfoil due to the pitching motion $\alpha_{\mathrm{geo}}(t)$, as shown in Fig.\ \ref{fig:AlphaEff} for the common baseline case with $\hat{\alpha}_{\mathrm{eff}} = 30 \, ^{\circ}$.
By adapting $\alpha_{\mathrm{geo}}(t)$ for different dimensionless parameters, $\hat{\alpha}_{\mathrm{eff}}$ can be kept constant. Additionally, different $\hat{\alpha}_{\mathrm{eff}}$ can be produced independent of the dimensionless parameters. All kinematics investigated in this study are designed with a quasi-sinusoidal evolution of $\alpha_{\mathrm{eff}}(t)$. As an extract from cyclic motion, only the downstroke of the airfoil is investigated.

The execution of combined pitching and plunging kinematics at TUDA was realized by attaching the flat plate to two linear actuators of type \textit{LinMot} PS01 - 48x240F - C with the aid of a midspan bracket, one at the leading edge and a second one at about $x/c=$ 0.68. By execution of different motion profiles on both actuators, pitching motion could be added to pure plunge motion. To produce an accurate motion profile and vibration free translation of the airfoil, the position and acceleration of the actuators were used by the actuator control as a feedback loop input.

A customized experimental platform, including a rotating stage and a linear translation stage, was designed to enable the combined pitching and plunging motion of the flat plate at BUAA. The linear translation of the investigated airfoil to produce plunging motion was realized with a servo motor of type \textit{YASKAWA} SGM7J-100 and a ball screw rod (LC-EA-030A). Rotational motion of the airfoil was realized with a servo motor of type \textit{YASKAWA} SGM7J-400 and a decelerator (\textit{KAMO} JFR90), which are both directly connected to the airfoil. A programmable multi-axis controller (\textit{Delta Tau} Clipper) was used to synchronize all the servomotors.

\begin{figure}[h]
    	\includegraphics[width=\columnwidth]{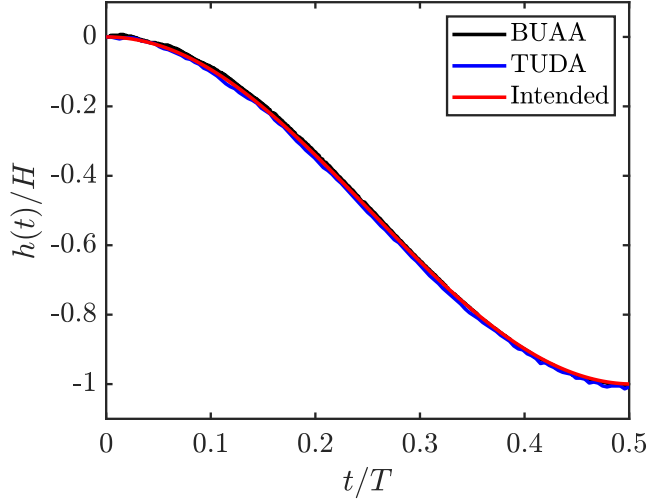}
        \caption{Comparison of the intended and experimentally realised vertical leading edge positions $h$, normalized by the plunge height $H$ as a function of  dimensionless time ($t/T$)  for the common baseline case.}
        \label{fig:LEPosition}
        \end{figure}

To allow flow fields to be compared between both facilities, the vertical leading edge position of the airfoil $h(t)$ was extracted from masked raw PIV images via image processing. In Fig.\ \ref{fig:LEPosition}, $h(t)$ of both facilities and the intended curve  are shown for the common baseline case, normalized with the respective full stroke height $H$.\\
Deviations between the measured and intended leading edge positions were found to be smaller than 0.6 \% of the full stroke height for all investigated cases, which is within the accuracy of the actuators used to pitch and plunge the flat plates. These deviations are small enough to allow velocity fields and vortex characteristics from both setups to be directly compared to one another.

After establishing good agreement between results from both facilities for the baseline case, a larger parameter range was tested at both facilities to characterise secondary structure emergence and their dynamics during vortex detachment. Although the dimensionless parameter range covered by this study is within the range of MAVs, it is also assumed to be transferable to biological propulsion as discussed and quantified in Sect.\ \ref{sec:sec_Intro}. The Reynolds number was fixed at 24,000 for all experiments while the reduced frequency was varied between 0.3 and 0.48 and the Strouhal number between 0.04 and 0.16. Table \ref{tab:AllCasesTUDA} lists all cases investigated at TUDA including the common baseline case (ID 3) with their dimensionless parameters and the geometric parameters $\hat{\alpha}_{\mathrm{eff}}$ and $\hat{\alpha}_{\mathrm{geo}}$.
    \begin{table}[!ht]
    \centering
    \begin{tabular}{| c | c | c | c | c | c |}
    \hline
    ID & Symbol & $k$ & $St$ & $\hat{\alpha}_{\mathrm{eff}}$ ($\, ^{\circ}$) & $\hat{\alpha}_{\mathrm{geo}}$ ($\, ^{\circ}$)\\ \hline\hline	
    1 & \includegraphics[scale=0.13, trim = 0 0 0 0, clip]{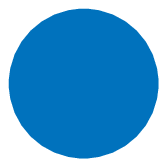} & 0.48 & 0.04 & 30 & 22.69 \\ \hline
    2 & \includegraphics[scale=0.13, trim = 0 0 0 0, clip]{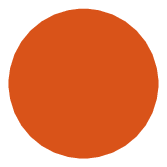} & 0.48 & 0.08 & 30 & 16.07 \\ \hline
    3 & \includegraphics[scale=0.13, trim = 0 0 0 0, clip]{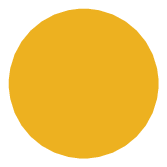} & 0.48 & 0.1 & 30 & 12.67 \\ \hline
    4 & \includegraphics[scale=0.13, trim = 0 0 0 0, clip]{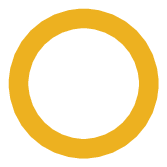} & 0.48 & 0.1 & 20 & 2.67 \\ \hline
    5 & \includegraphics[scale=0.13, trim = 0 0 0 0, clip]{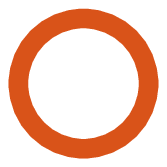} & 0.48 & 0.08 & 20 & 6.07 \\ \hline
    6 & \includegraphics[scale=0.13, trim = 0 0 0 0, clip]{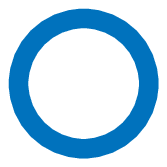} & 0.48 & 0.04 & 20 & 12.7 \\ \hline
    7 & \includegraphics[scale=0.13, trim = 0 0 0 0, clip]{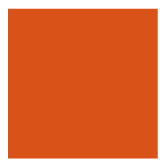} & 0.3 & 0.08 & 30 & 15.95 \\ \hline
    8 & \includegraphics[scale=0.13, trim = 0 0 0 0, clip]{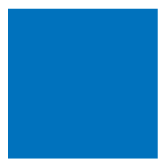} & 0.3 & 0.04 & 30 & 22.86 \\ \hline
    9 & \includegraphics[scale=0.13, trim = 0 0 0 0, clip]{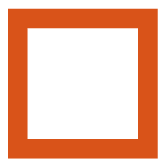} & 0.3 & 0.08 & 20 & 5.95 \\ \hline
    10 & \includegraphics[scale=0.13, trim = 0 0 0 0, clip]{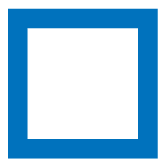} & 0.3 & 0.04 & 20 & 12.86 \\ \hline
    \end{tabular}
    \caption{Dimensionless and geometric parameter range investigated at TUDA for $Re = 24,000$, including the common baseline case (ID 3).}
    \label{tab:AllCasesTUDA}
    \end{table}
The ID assigned to each case will be used for identification in subsequent evaluations. The $St$-range at TUDA was limited to below 0.1 due to high inertial forces of fast motion kinematics in air. The difference in dynamic motions between facilities becomes evident by considering the motion period for the common baseline case. In air the downstroke has to be completed within 0.118\,s while in water it lasts 3.927\,s. By including a set of cases with $\hat{\alpha}_{\mathrm{eff}}= 20 \, ^{\circ}$ (ID 4 to 6 and 9 to 10) and variations of $St$ (ID 1 to 3) and $k$ (ID 7 and 8) with respect to the baseline case, effective inflow angles close to pure plunging motion, indicated by their low $\hat{\alpha}_{\mathrm{geo}}$, can also be investigated. The parameter space investigated at BUAA is depicted in table \ref{tab:AllCasesBUAA}, again including the common baseline case referred to ID3 (TUDA) and ID 11 (BUAA).   
   \begin{table}[!ht]
    \centering
    \begin{tabular}{| c | c | c | c | c | c |}
    \hline
    ID & Symbol & $k$ & $St$ & $\hat{\alpha}_{\mathrm{eff}}$ ($\, ^{\circ}$) & $\hat{\alpha}_{\mathrm{geo}}$ ($\, ^{\circ}$)\\ \hline\hline
    11 & \includegraphics[scale=0.13, trim = 0 0 0 0, clip]{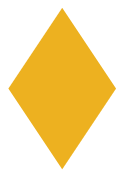} & 0.48 & 0.1 & 30 & 12.67 \\ \hline
    12 & \includegraphics[scale=0.13, trim = 0 0 0 0, clip]{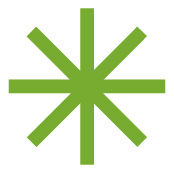} & 0.48 & 0.16 & 30 & 3.31 \\ \hline
    13 & \includegraphics[scale=0.13, trim = 0 0 0 0, clip]{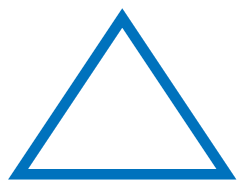} & 0.3 & 0.04 & 20 & 12.86 \\ \hline
    \end{tabular}
    \caption{Dimensionless and geometric parameter range investigated at BUAA for $Re = 24,000$, including the common baseline case (ID 11).}
    \label{tab:AllCasesBUAA}
    \end{table}
With the higher attainable Strouhal number the dimensionless parameter range at BUAA could be extended towards the regime of efficient forward flight, characterized by higher $St$.

At TUDA raw PIV images were acquired with a double frame recording frequency of 1 kHz and an inter-framing time of $\triangle t = 150$\,$\mu$s, yielding a particle displacement between 2 and 8 pixels for PIV correlations. DEHS seeding particles with a mean diameter of 0.5 - 1\,$\mu$m were introduced into the settling chamber of the wind tunnel. Their response time $\tau_{s}$ was computed to be 2.7\,$\mu$s according to \cite{Raffel.2007}. Compared to the minimum time scale of the flow, estimated by the convective time of the Kolmogorov length scale $\eta_{K}=c \times R e^{-3 / 4} \approx 60 \, \mu \mathrm{m}$ with $c$ as the macro-scale reference, $\tau_{F}=\eta_{K} / U_{\infty}$ = 20\,$\mu$s, the response time is considered to be adequate. A \textit{Photron SA1.1} High Speed Camera and a \textit{Carl Zeiss} lens of $50$\,mm focal distance with an aperture of $f=$ 2.2 captured images of the seeded flow around the flat plate with a resolution of 1024 $\times$ 1024 pixels. The field of view spanned by a light sheet plane of 3\,mm thickness, located at 28\% span from the wind tunnel wall, was 1.9$c$ $\times$ 1.9$c$, resulting in a spatial resolution of 4.556\,px/mm. The seeded flow field was illuminated using a dual cavity \textit{Litron} DY-303 Nd:YLF laser with a single pulse energy of 18\,mJ. Image correlations were perfomed with \textit{PIVview} 2C software from \textit{PIVTEC GmbH} utilizing a multi-grid, multi-pass interrogation scheme including a sub-pixel routine at an initial interrogation area (IA) size of 64\,px\,$\times$\,64\,px and a final of 12\,px\,$\times$\,12\,px at 50\,\% overlap, yielding 92 IA's over the chord. Outliers were identified by a median test \citep[cf.][]{Westerweel.2005} in a 3\,$\times$\,3 neighbourhood with a threshold of twice the velocity magnitude and found to be always less than 1.6 \% of all computed vectors for all acquired image pairs. Each parameter set was recorded 10 times yielding a spatially averaged standard deviation of flow fields within 3.1\,\% and a temporally averaged deviation of vortex characteristics of 2.1\,\% of the respective asymptotic values. Standard deviation values were computed by using bootstrap testing with 1000 bootstrap samples for each investigated number of repetitions \citep[cf.][]{Benedict.1996}.

In the BUAA setup a high-speed CMOS camera of type \textit{Photron Fastcam} SA2/86K-M3 fitted with a Nikon lens of 50\,mm focal length was used to acquire time-resolved PIV raw images. Image pairs were acquired at a frequency of 200\,Hz with a resolution of 2048\,px $\times$ 2048\,px. The seeded flow field was illuminated by a continuous Nd-YAG laser with 8\,W nominal power at midspan position with a laser light sheet of 3\,mm thickness. Hollow glass beads with a median diameter of 20\,$\mu$m and a density of 1.05\,g/cm$^{3}$ were used as seeding particles in water. For final evaluations, the same correlation algorithm from TUDA was used, where the final IA size was 16\,px $\times$ 16\,px with 50\% overlap resulting in about 82 velocity data points over the airfoil chord.

\subsection{Data Processing}
\label{SubSec:DataProc}
The circulation and position of the LEV are characteristic parameters that can be compared between both setups and used for further evaluations. To investigate influences of the evaluation method used to obtain circulation and position, different methods have been tested on the same set of raw images. Investigated methodologies to obtain LEV circulation are based on the identification of regions belonging to the vortex prior to a first order spatial integration of vorticity according to Stokes's theorem. Vortex characteristics were extracted from single runs before they were ensemble averaged.

LEV boundaries computed using the $\lambda_{ci}$ method by \cite{ZHOU.1999} were found to be strongly dependent on the threshold used to identify the vortex, whereas the Q criterion by \cite{Hunt.1988} identified only an inner vortex core while excluding outer vortex regions. Additionally, both methods were found to intermittently attribute the leading edge shear layer to the main vortex, leading to strong fluctuations of the subsequently determined circulation. The LEV boundary computed by thresholding the $\Gamma_{\mathrm{2}}$ scalar field, introduced by \cite{Graftieaux.2001}, which considers regions of pure shear as the vortex area, was found to identify the vortex boundary most consistently for both data-sets using the default threshold of $\Gamma_{\mathrm{2}} = 2/\pi$. Therefore, it was used to quantify the circulation evolution of the LEV from velocity fields obtained at both facilities. This was done by integrating vorticity within the detected LEV boundary according to Stokes's theorem. The detection of the LEV center from maxima of the $\Gamma_{1}$ scalar function, also introduced by \cite{Graftieaux.2001}, as well as from the Q criterion, were found to be reliable, at least with respect to visual inspection of instantaneous vector fields. Although the $\Gamma_{\mathrm{1}}$ function is not Galilean invariant by definition, deviations from the Galilean invariant Q criterion were found to be less than 2\%, which in turn provides evidence that the chosen plate-fixed frame of reference allows reasonable interpretation of the extracted topology.

Even when using the same vortex boundary identification and circulation computation method, as well as spatial vorticity derivation schemes, the LEV circulation evaluation was found to be dependent on the cross-correlation parameters used to obtain velocity data from the same raw images in PIV correlations. The LEV circulation obtained with identical post-processing routines but different correlation algorithms, a gradient-based cross-correlation optimization based on the Lucas-Kanade method \citep[cf.][]{Champagnat.2011} and a standard FFT correlation algorithm, differs up to 9\,\%. The maximum circulation divergence occurs when the LEV starts to decay, which is accompanied by a blurred outer boundary leading to different areas identified as a vortex. This highlights that PIV correlation schemes influence the circulation computation within the vortex domain significantly, even when using the same vortex identification method for boundary detection, vorticity derivation and subsequent spatial integration procedures. When aiming to compare vortex characteristics between different facilities, these deviations can hinder comparability of characteristics obtained from flow fields. Consequently all evaluations have been performed using the same correlation software and parameters according to the TUDA setup described in section\ \ref{SubSec:FacPar}. Data sets in terms of correlated velocity fields of the baseline cases from TUDA and BUAA (ID3 and ID 11) are available online as reference cases (http://dx.doi.org/10.25534/tudatalib-168).

With Eulerian vortex identification methods the LEV detachment, related to changes of the flow topology, can only be investigated implicitly by observation of effects on vortex characteristics. In contrast, Lagrangian methods in terms of coherent structures allow  direct identification of topological changes of the flow field initiating vortex detachment. In this study coherent structures are identified with the aid of finite-time Lyapunov exponent (FTLE) ridges, following the concept and methods introduced by \citet{Haller.2002} and \citet{Shadden.2006}. Ridges of repelling and attracting fluid regions are obtained by thresholding forward-time and backward-time FTLE scalar fields. Topological changes of the flow field can be identified directly by tracking Lagrangian saddle points of the flow field, which are intersection points of forward-time and backward-time FTLE ridges, as shown by \cite{Huang.2015}. In this study the FTLE computation package developed by \citet{Peng.2009} is used to obtain FTLE fields.

In advance of FTLE evaluations, the impact of different parameters governing the results of FTLE computations, including spatial and temporal resolution of the velocity information as well as the integration time $\tau$ were investigated. Computation of physically relevant FTLE ridges requires a sufficient spatial and temporal resolution of the velocity information used for calculations. Both resolutions were tested for current FTLE computations by a reduction of the temporal and spatial resolution by a factor of 2. In both cases no significant changes were observed in FTLE fields, so both resolutions are assumed to be sufficient. FTLE scalar fields are obtained over a time frame, which is referred to as the integration time $\tau$. As mentioned by \citet{Peng.2009} $\tau$ does not have an impact on the FTLE ridge topology in terms of their location but on the strength of ridges via their resolution. When thresholding ridges from raw FTLE fields using a percentage of maximum FTLE values in each frame to evaluate distinct ridges (80\,\% in this study), the strength of ridges can determine whether they pass the threshold. The final integration time of $\tau = 0.11$ $T/t$  was chosen, since ridges obtained using the 80\,\% threshold represented all significant features of the raw contours, while the computational effort remained acceptable. Due to the robustness of FTLE calculations against interpolation errors \citep[cf.][]{Haller.2002}, velocity vector fields can be interpolated to obtain highly resolved FTLE fields. In this study the spatial interpolation is chosen based on the trade-off between resolution and computational costs. In final evaluations the spatial resolution of FTLE fields was twice as high as the resolution of the velocity vector fields, since ridges were clearly represented and computational costs were acceptable.

\section{Results}
With the aim of comparing results  from the two different facilities at BUAA and TUDA in water and air, this section focuses on the evaluation of the common baseline case. 

\subsection{Flow Fields}
\begin{figure*}[!ht]
    \begin{subfigure}[t]{0.259\textwidth}
        \centering
        \includegraphics[width=\textwidth, trim = 10 0 140 0, clip]{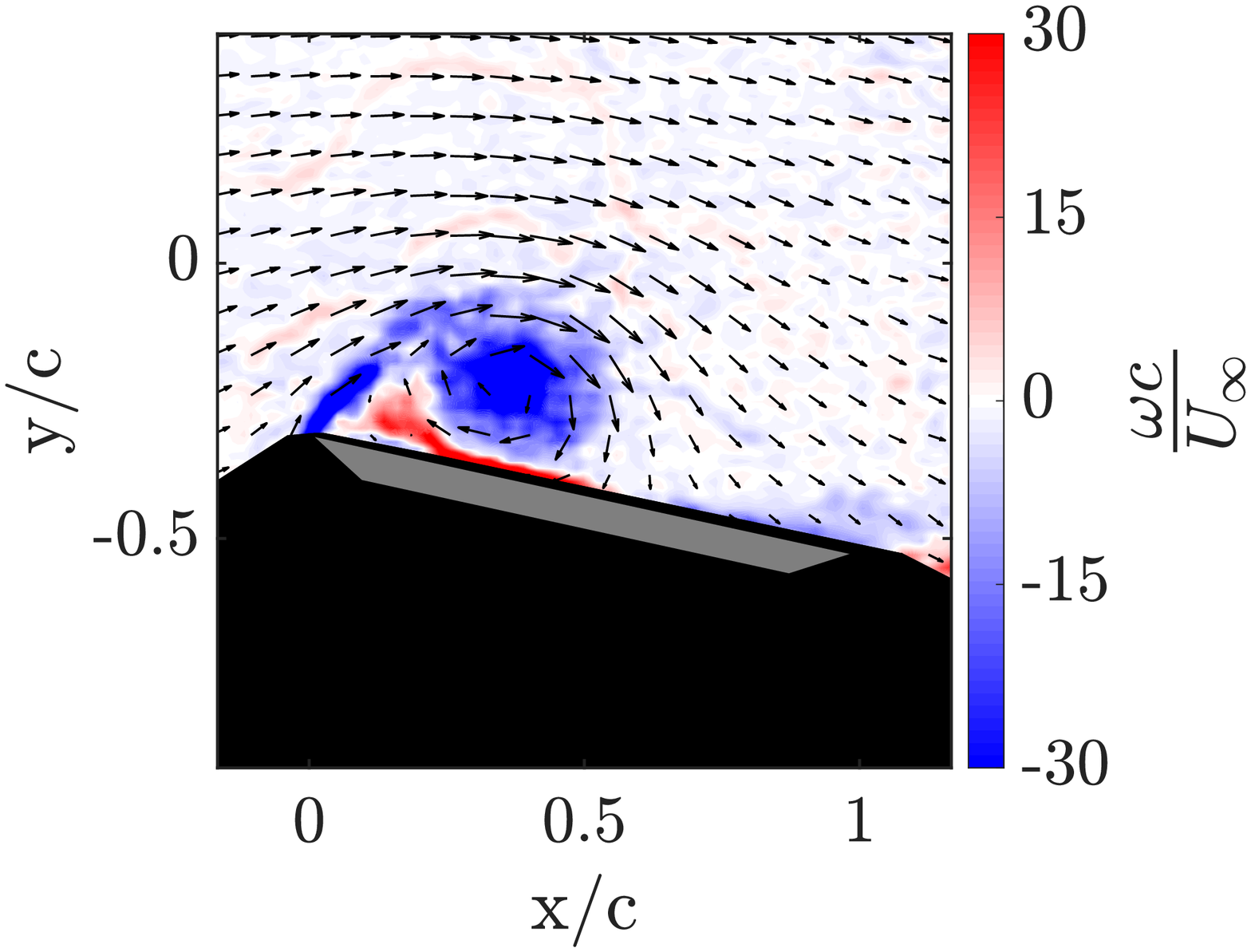}
        \caption{BUAA; $t/T = 0.25$}
        \label{fig:Vort_BUAAtT0x15}
        \end{subfigure}
        \begin{subfigure}[t]{0.199\textwidth}
        \centering
        \includegraphics[width=\textwidth, trim = 115 0 140 0, clip]{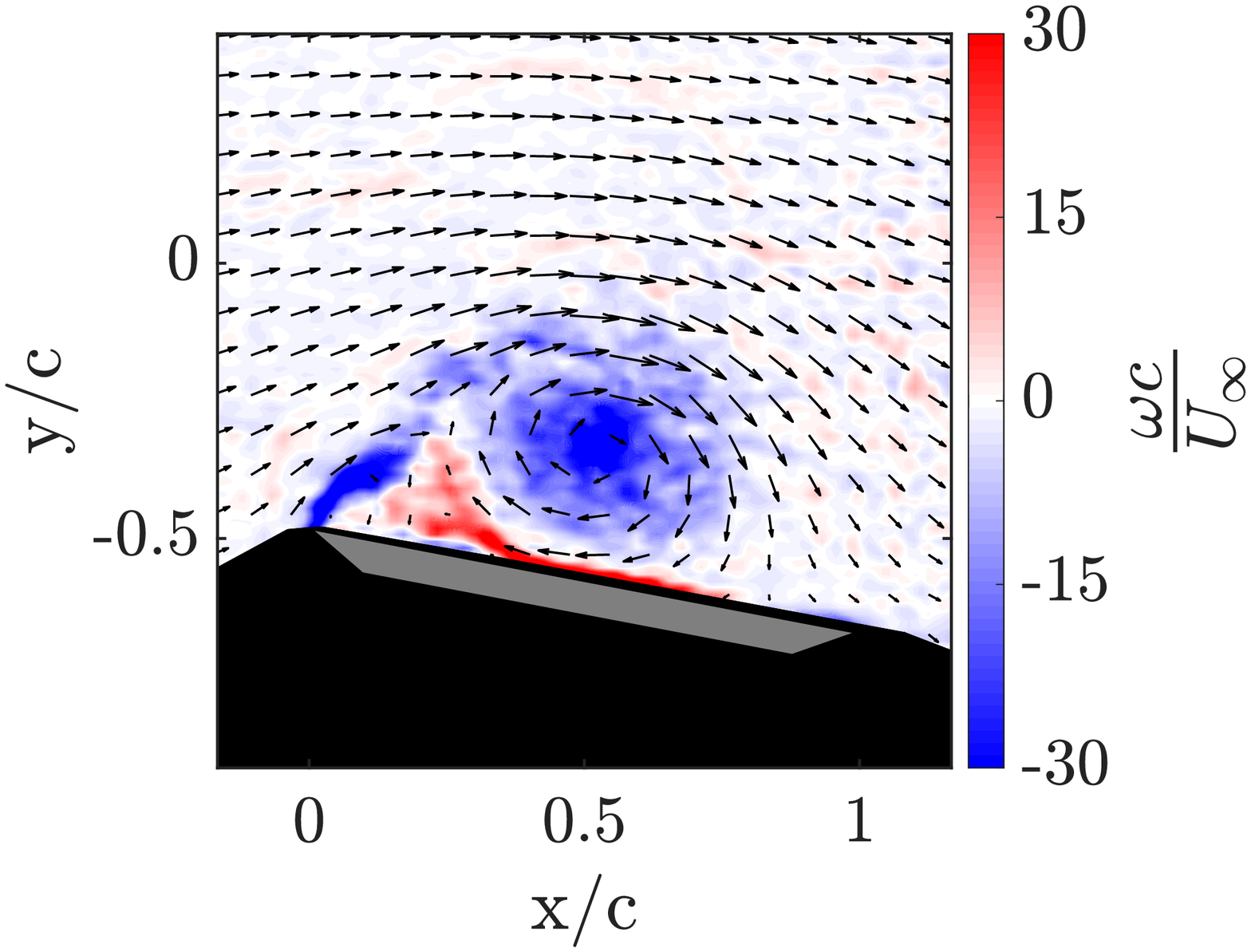}
        \caption{BUAA; $t/T = 0.35$}
        \label{fig:Vort_BUAAtT0x25}
        \end{subfigure}
        \begin{subfigure}[t]{0.199\textwidth}
        \centering
        \includegraphics[width=\textwidth, trim = 115 0 140 0, clip]{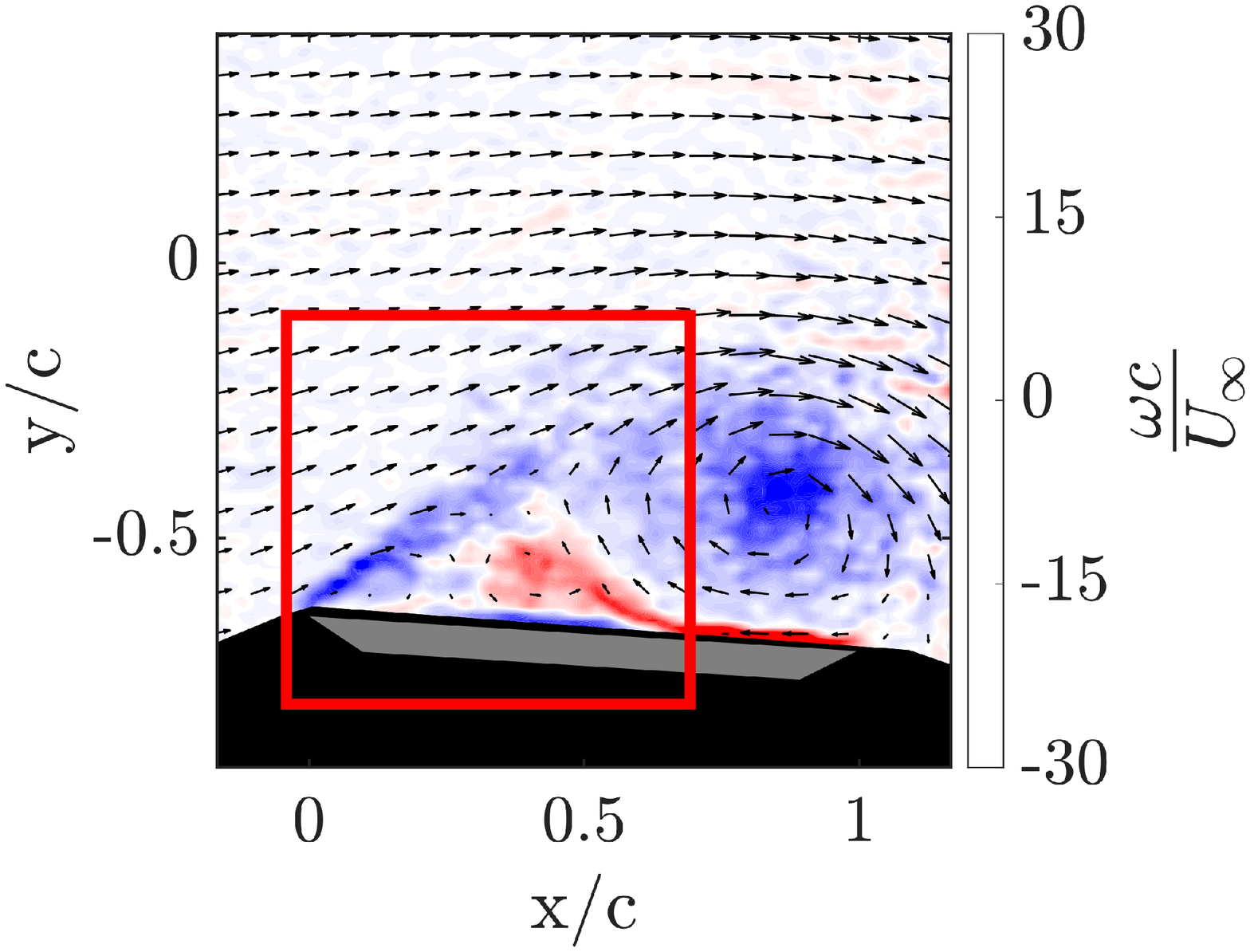}
        \caption{BUAA; $t/T = 0.45$}
        \label{fig:Vort_BUAA_tT0x35}
        \end{subfigure}
        \begin{subfigure}[t]{0.315\textwidth}
        \centering
        \includegraphics[width=\textwidth, trim = 50 0 5 0, clip]{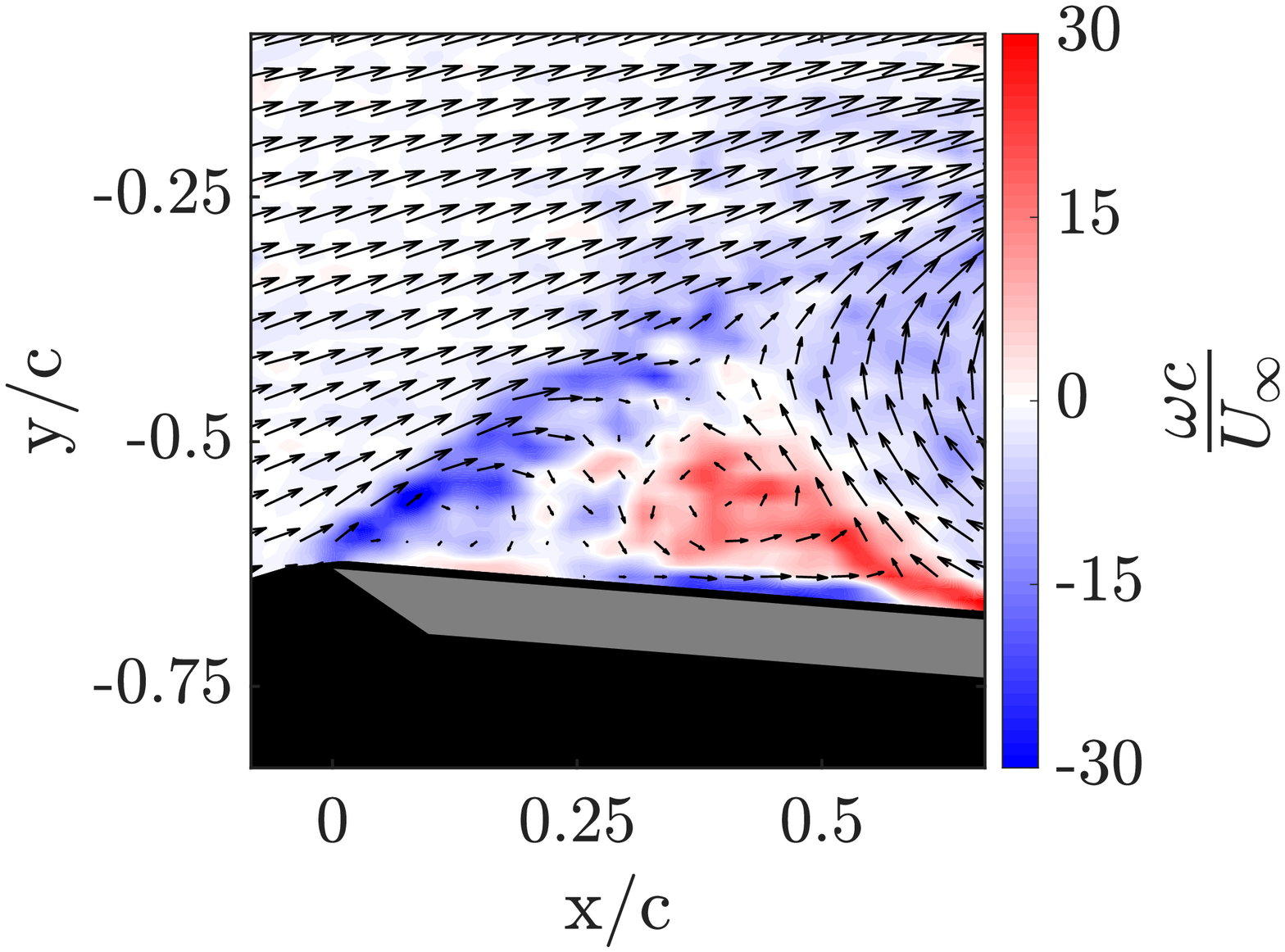}
        \caption{BUAA; $t/T = 0.45$}
        \label{fig:Vort_BUAA_tT0x5}
        \end{subfigure}
	    \begin{subfigure}[t]{0.259\textwidth}
        \centering
        \includegraphics[width=\textwidth, trim = 10 0 140 0, clip]{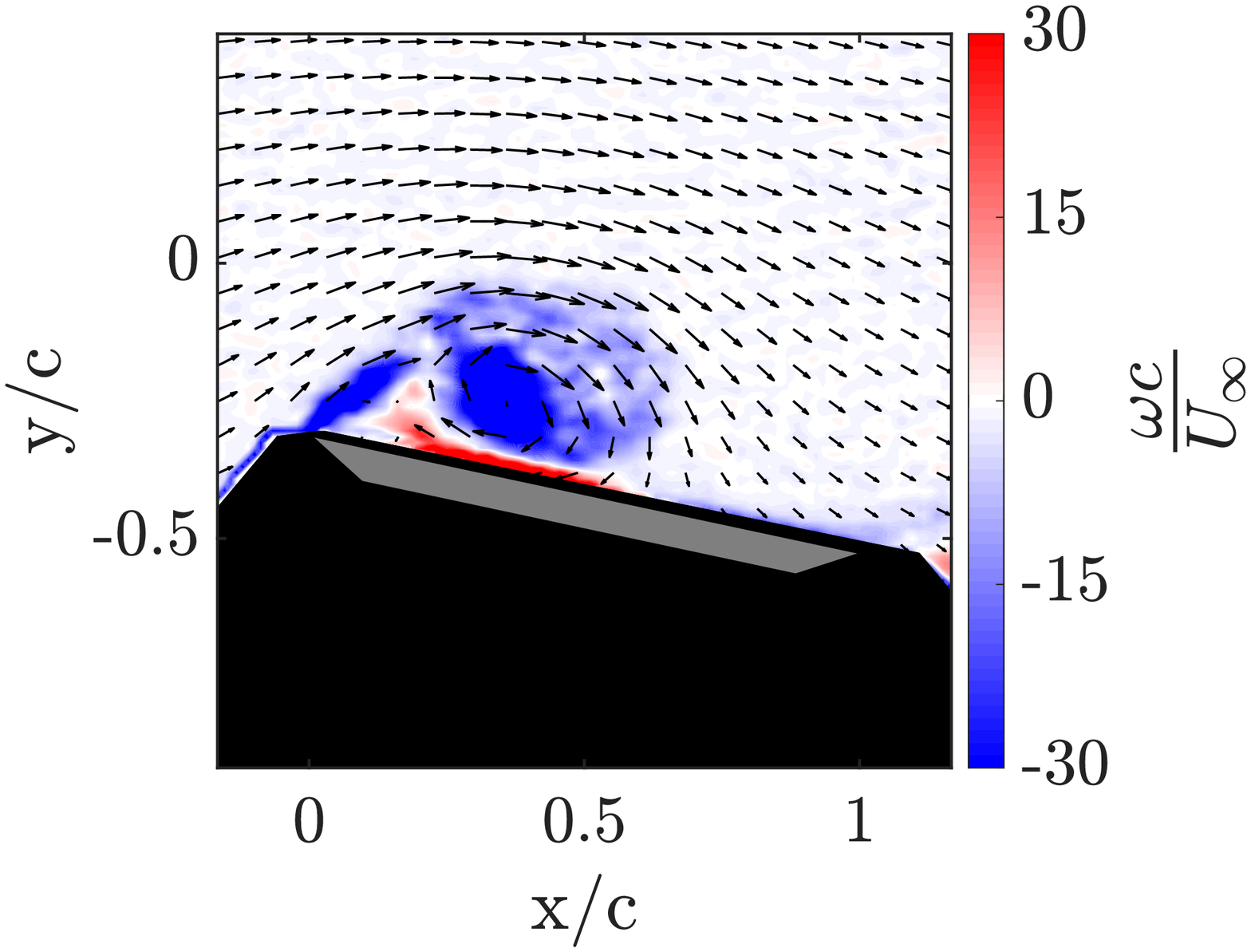}
        \caption{TUDA; $t/T = 0.25$}
        \label{fig:Vort_TUDAtT0x15}
        \end{subfigure}
        \begin{subfigure}[t]{0.199\textwidth}
        \centering
        \includegraphics[width=\textwidth, trim = 115 0 140 0, clip]{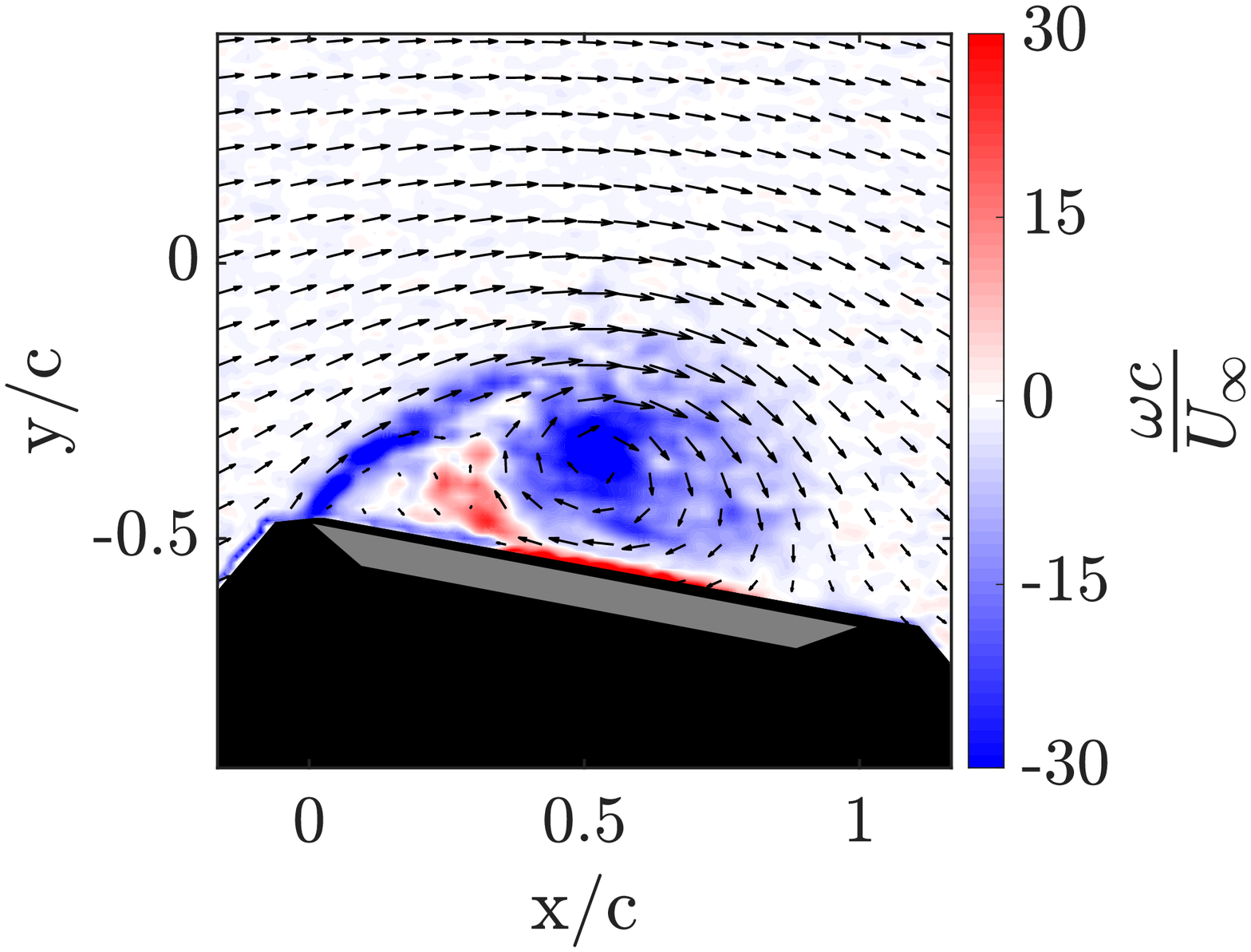}
        \caption{TUDA; $t/T = 0.35$}
        \label{fig:Vort_TUDAtT0x25}
        \end{subfigure}
        \begin{subfigure}[t]{0.199\textwidth}
        \centering
        \includegraphics[width=\textwidth, trim = 115 0 140 0, clip]{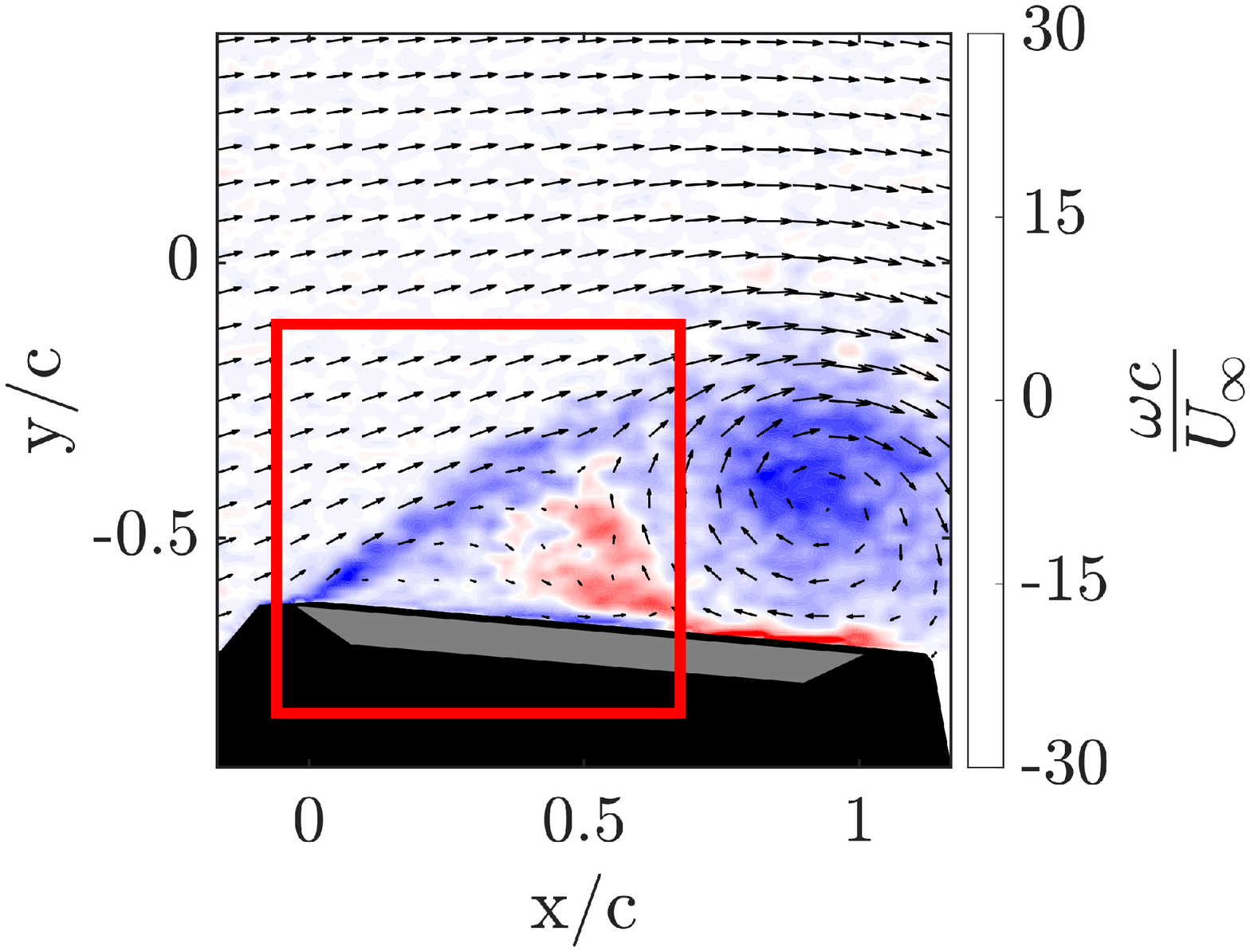}
        \caption{TUDA; $t/T = 0.45$}
        \label{fig:Vort_TUDA_tT0x35}
        \end{subfigure}
        \begin{subfigure}[t]{0.315\textwidth}
        \centering
        \includegraphics[width=\textwidth, trim = 50 0 5 0, clip]{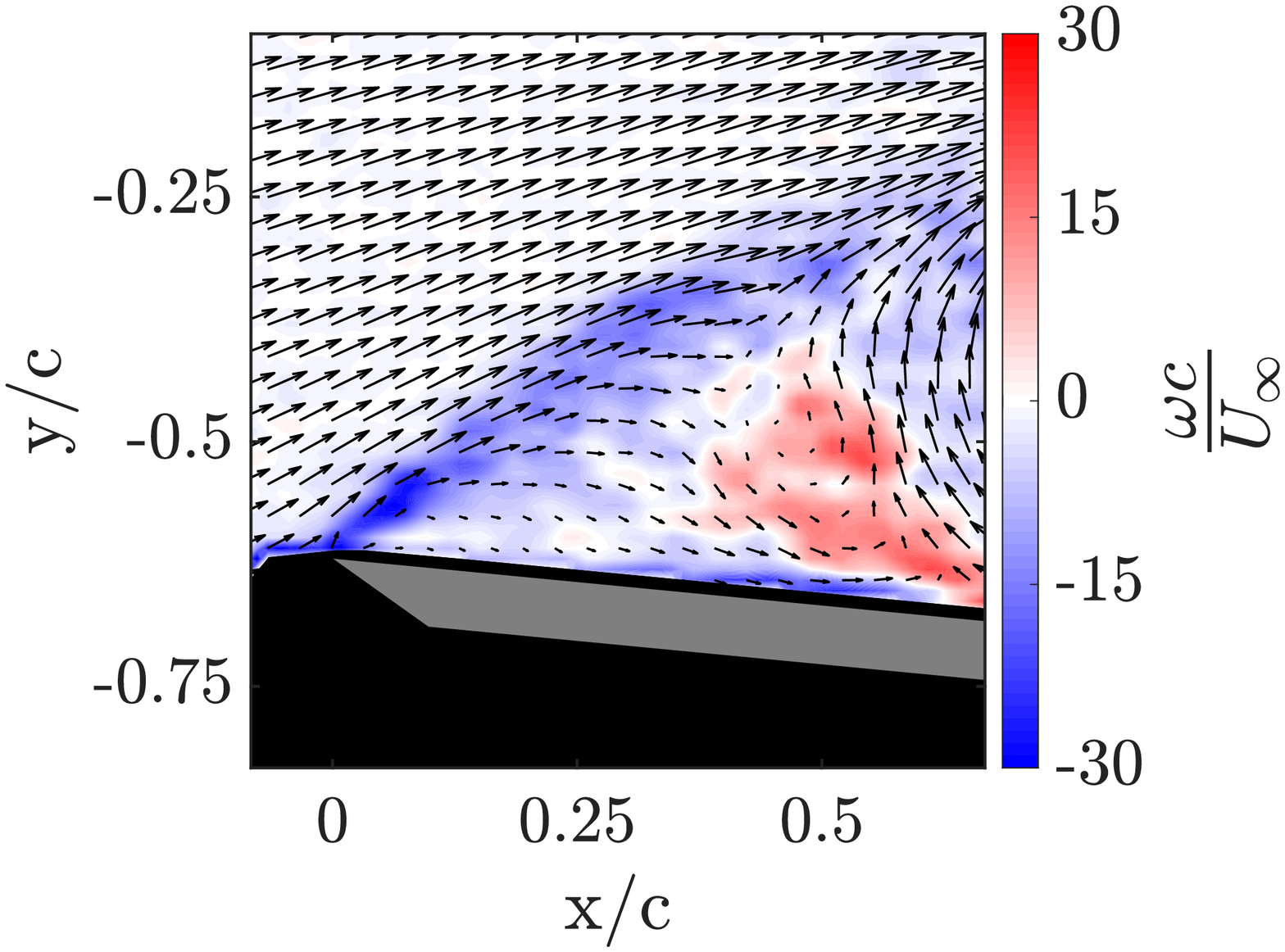}
        \caption{TUDA; $t/T = 0.45$}
        \label{fig:Vort_TUDA_tT0x5}
        \end{subfigure}
    \caption{Flow field in terms of normalized vorticity $\omega c/ U_{\infty}$ for different dimensionless time instants $t/T$ from ensemble averaged PIV measurements at BUAA ((a)-(d)) and TUDA ((e)-(h)) for the common baseline case (ID 11 case from BUAA and ID 3 from TUDA). The inflow is from the left, the airfoil is masked out in grey and the laser light shadow caused by the airfoil in black. While (a)-(c) and (e)-(g) depict the entire field with only every 6th vector for clarity, (d) and (h) show the leading edge region with every 3rd vector as highlighted in (c) and (g).}
    \label{fig:Vort}
    \end{figure*}

A qualitative aspect of comparability is the evolution of the flow field and its topology, which is depicted in terms of ensemble averaged vorticity fields in Fig.\ \ref{fig:Vort}. 
The coordinate system is fixed at the initial top position of the airfoil leading edge and both axes are normalized by the flat plate chord. At $t/T=$ 0.25 the LEV containing concentrated negative (blue color coded) vorticity grows on the airfoil by accumulating circulation from the leading edge shear-layer (Figs.\ \ref{fig:Vort} (a) and (e)). A thin layer of counterclockwise rotational fluid, indicated by the positively signed (red color coded) vorticity, can be observed below the vortex in addition to a distinct region comprising positively signed vorticity ahead of the main vortex. This counterclockwise rotational fluid forms a secondary vortex rotating in the opposite direction of the clockwise rotating LEV. A third clockwise rotating vortex between the leading edge shear layer and the secondary vortex completes the secondary vortex structures. This is identifiable from the indicated velocity vectors in Figs.\ \ref{fig:Vort} (d) and (h) later in the downstroke. At this early stage the flow fields from both setups are in very good agreement with respect to the observed topology and vorticity intensity within the vortices.

Furthermore, towards the end of the downstroke, the LEV continues growing, although at $t/T=$ 0.35 the connection between vortex and leading edge shear layer appears to be interrupted in the vorticity fields from BUAA (Fig.\ \ref{fig:Vort} (b)), while it is unimpaired in the TUDA fields (Fig.\ \ref{fig:Vort} (f)). An inspection of the entire time series of the flow fields revealed that the LEV and shear layer are frequently interrupted for TUDA results, while for BUAA flow fields their connection appears to be generally weaker, indicated by lower vorticity values of the connecting region. Differences of single events in the instantaneous velocity fields are of minor relevance for the present study, aimed at highlighting the main features of the ensemble averaged velocity fields. Shortly before the end of the downstroke at $t/T=$ 0.45, a slight deviation of the LEV center position can be observed comparing Figs.\ \ref{fig:Vort} (c) and (g). At this late stage of the downstroke the LEV is only weakly connected to the leading edge shear layer for both evaluations, indicating the end of circulation accumulation. Overall, the topological evolution of the ensemble averaged flow fields and the qualitative vortex characteristics evolution are in very good agreement between the two facilities.

\subsection{Vortex Characteristics}
\begin{figure}[ht]
    	\includegraphics[width=\columnwidth, trim = 0 5 25 15 ,clip]{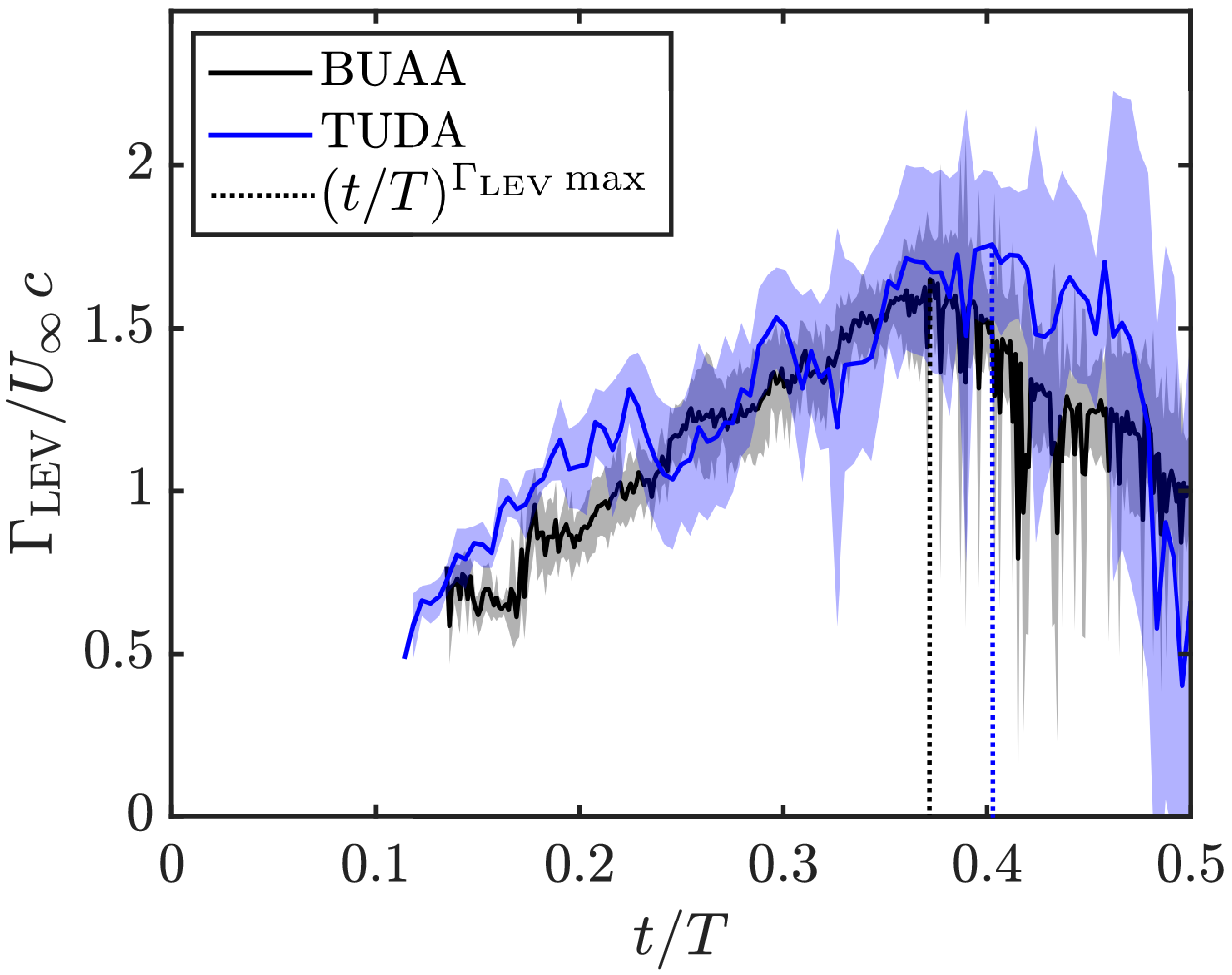}
        \caption{Evolution of the normalized leading edge vortex circulation $\Gamma_{\mathrm{LEV}} / U_{\infty} \, c$ for the common baseline case from both setups. The respective standard deviation is indicated as a coloured shadow and the peak circulation $(t/T)^{\mathrm{\Gamma_{LEV}\,max}}$ as dashed line. Circulation is obtained by integration of vorticity over the vortex area identified from the $\Gamma_{2}$ scalar field.}
        \label{fig:CircBase_Gamma2}
    \end{figure}

The LEV characteristics in terms of its position, size and circulation determine the lift that is induced on the pitching and plunging flat plate. Therefore, these quantities are of key interest in future flow control approaches. To allow comparability between different flow control approaches at both facilities, vortex characteristics of the common baseline case will now be compared quantitatively. Fig.\ \ref{fig:CircBase_Gamma2} depicts the normalized LEV circulation $\Gamma_{\mathrm{LEV}} / U_{\infty} c$ determined by the integration of vorticity within its detected boundary of the $\Gamma_{2}$ scalar field. Raw images were correlated using the same algorithm implemented at TUDA and the same vortex identification method in addition to the same vorticity derivation and integration schemes. 
The circulation of the vortex was computed for each individual run prior to ensemble averaging of the extracted circulation for all runs. In this manner the standard deviation can be computed for the final, ensemble averaged results, indicated as coloured shadow in Fig.\ \ref{fig:CircBase_Gamma2}. Early during vortex growth, the LEV circulation increases with a constant offset, leading to higher vortex circulation from TUDA results up to $t/T\approx$ 0.22. After this phase, stronger normalized circulation fluctuations are observed in the TUDA results compared to those from BUAA. These fluctuations and the initial offset originate from an intermittent inclusion of the leading edge shear layer into the computational vortex boundary, evident in single frame evaluations, which are only observed for TUDA experiments. This was confirmed by simultaneous fluctuations of the detected area of the vortex boundary. The LEV stops accumulating circulation between 0.38 $\le t/T \le$ 0.41 in BUAA and TUDA results, indicated by the peak circulation instants $(t/T)^{\mathrm{\Gamma_{LEV}\,max}}$ (dashed lines in Fig.\ \ref{fig:CircBase_Gamma2}).

\begin{figure}[ht]
    	\includegraphics[width=\columnwidth, trim = 0 5 25 15 ,clip]{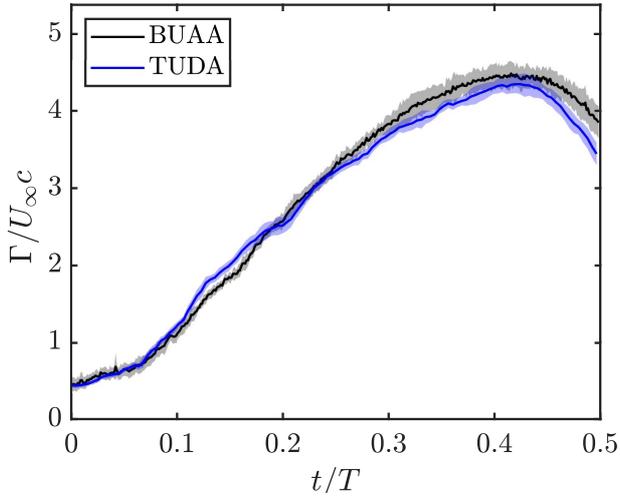}
        \caption{Evolution of the normalized circulation $\Gamma / U_{\infty} \, c$ for the common baseline case evaluated from the entire field of view to identify vortex boundary detection as the error source for circulation deviations. The standard deviation is indicated as  coloured shadows.}
        \label{fig:CircBase_FullFOV}
    \end{figure}

To verify that intermittent inclusion of the leading edge shear layer into the calculation of the circulation is an error source for deviations of the circulation evolution during the early LEV growth phase, the circulation was also computed using the entire field of view as an integration domain. A comparison of the normalized circulation from the entire field of view for both setups is shown in Fig.\ \ref{fig:CircBase_FullFOV}.
Despite very small deviations of the circulation, the overall quantitative agreement of the evolution is very good. In both experiments (at BUAA and TUDA), the maximum circulation is reached between 0.4151 $\le t/T \le$ 0.4195 with 3\% amplitude difference. Deviations in circulation remain below 5.3\% at any instant throughout the downstroke. Based on this good quantitative agreement it can be concluded that the observed deviations of circulation evolution in Fig.\ \ref{fig:CircBase_Gamma2} are caused by an intermittent inclusion of the leading edge shear layer into the integration domain for TUDA results during vortex identification. Potential reasons for these deviations could be the different spatial resolution of velocity fields or different free-stream turbulence levels of the tunnels used.

\begin{figure}[ht]
    	\includegraphics[width=\columnwidth, trim = 0 5 25 15 ,clip]{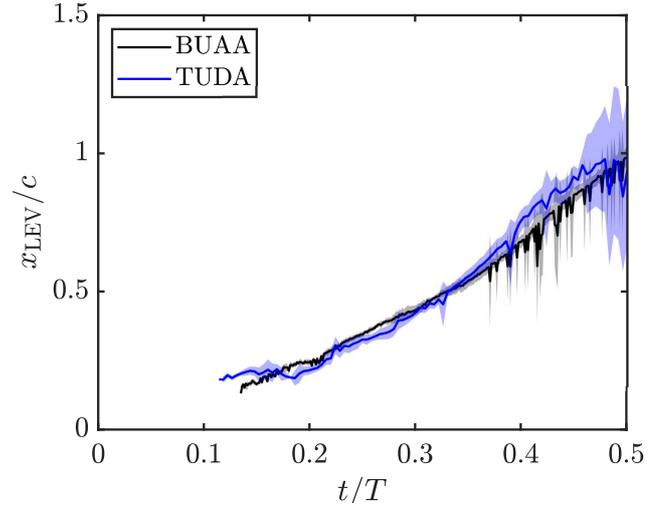}
        \caption{Evolution of the normalized streamwise LEV center position $x_{\mathrm{LEV}}/c$ in a plate-fixed frame of reference from the maximum of the $\Gamma_{1}$ scalar field for the common baseline case. The standard deviation is indicated as a coloured shadow.}
        \label{fig:XPositionBase}
    \end{figure}

Figure\ \ref{fig:XPositionBase} shows a comparison of the normalized streamwise LEV center position in a plate-fixed frame of reference.
Again, the position is extracted from phase-locked velocity fields taken at each individual run prior to ensemble averaging. The evolution is in good agreement up to $t/T=$ 0.39, whereupon small deviations of the position between BUAA and TUDA results occur. Since the accumulation of circulation in the vortex was found to stop at $t/T\approx$ 0.4, these deviations are attributed to the final convection of the vortex downstream of the airfoil and thus are not of interest for the current study.

Overall, good agreement of vortex characteristics is achieved between ensemble averaged results derived from the experiments performed at BUAA and TUDA.

\subsection{Detachment Mechanism}
Two different mechanisms of LEV detachment are postulated in literature, as pointed out in Sect.\ \ref{sec:sec_Intro}. One is related to fluid recirculation around the trailing edge of the airfoil, when the rear stagnation point of the flow behind the LEV on the airfoil convects beyond the trailing edge. This mechanism is investigated in the common baseline case of this study. As demonstrated by \cite{Rival.2014}, the extraction of the velocity immediately above and parallel to the airfoil surface allows the convection of the rear confining stagnation point of the LEV (marked as green diamond in Fig.\ \ref{fig:Topology}) to be tracked according to the change of velocity sign. The tangential velocity induced by vortices on the airfoil surface is schematically shown in Fig.\ \ref{fig:Topology} for an arbitrary point during airfoil motion. By identification of the rear stagnation point evolution, the temporal instant of maximum circulation can be compared to that of recirculation.

\begin{figure}[ht]
    \begin{subfigure}[t]{\columnwidth}
        \centering
        \includegraphics[width=\textwidth, trim = 0 0 0 0, clip]{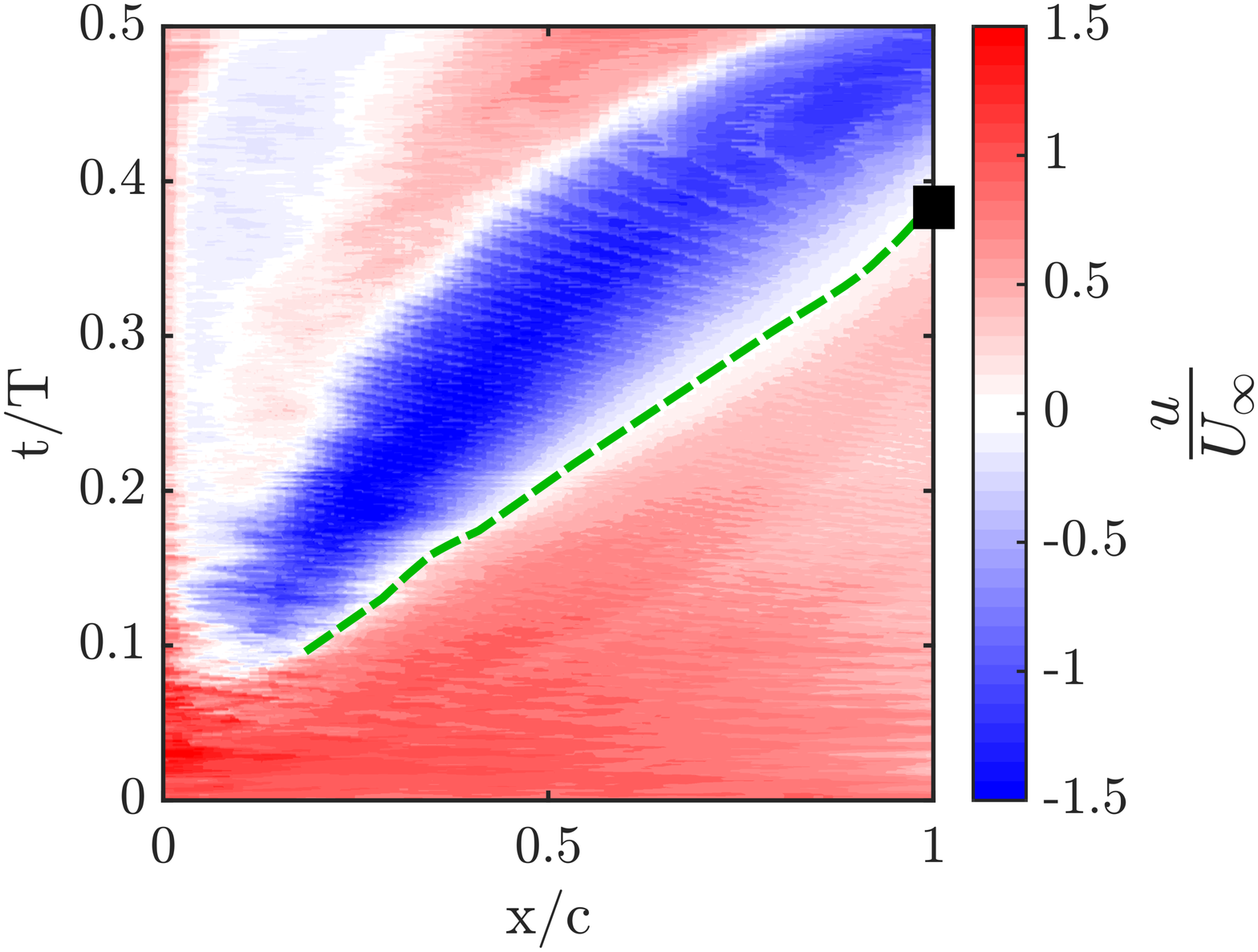}
        \caption{BUAA}
        \label{fig:Recirc_BUAA}
    \end{subfigure}
    \par\bigskip
	\begin{subfigure}[t]{\columnwidth}
        \centering
        \includegraphics[width=\textwidth, trim = 0 0 0 0, clip]{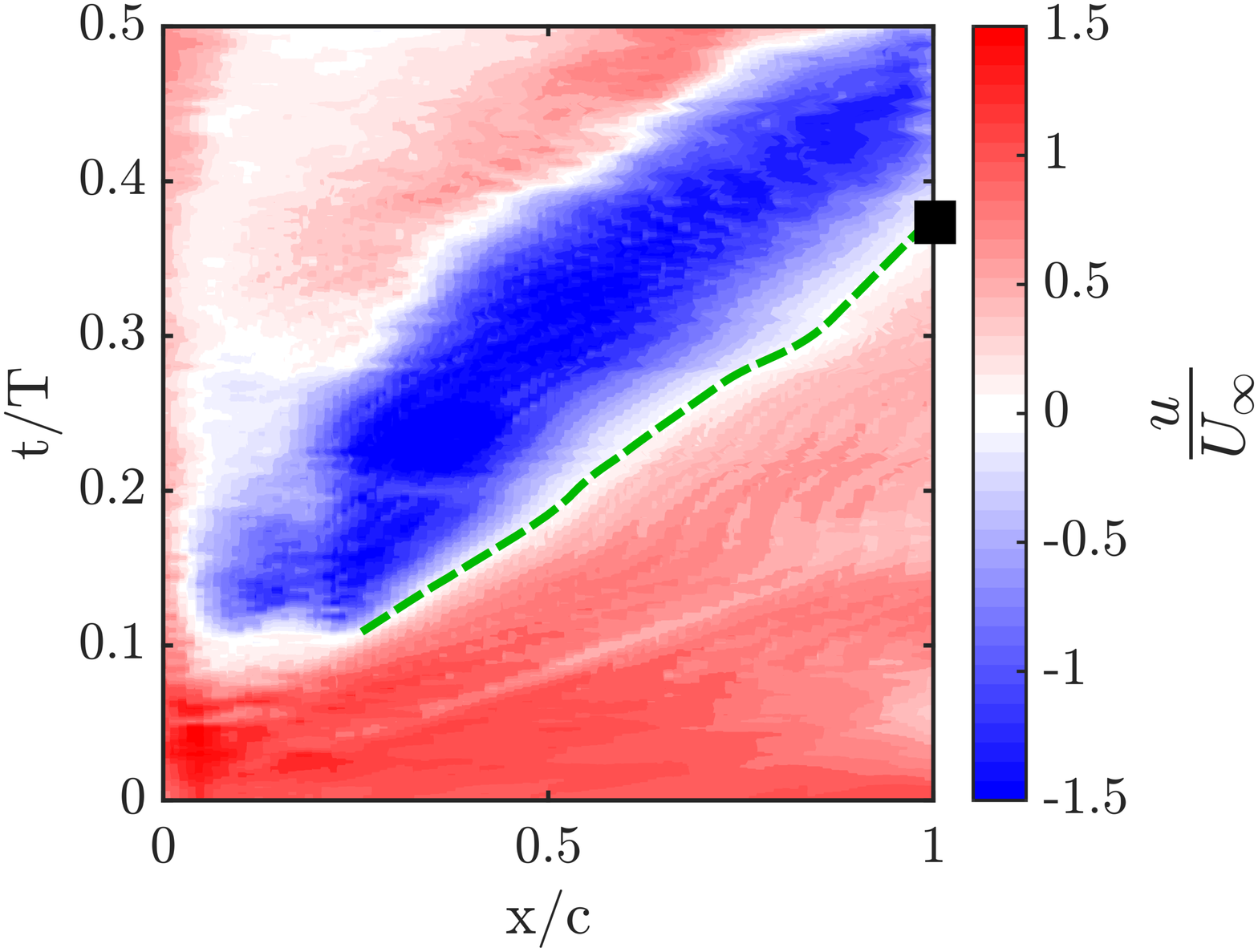}
        \caption{TUDA}
        \label{fig:Recirc_TUDA}
    \end{subfigure}
    \caption{Evolution of the velocity tangential to the airfoil surface $u$ scaled by the free-stream velocity $U_\infty$ over dimensionless chordwise position and time. The trace of the rear stagnation point of the flow behind the LEV is marked with a green dashed line; (a) tangential velocity from measurements in water at BUAA; (b) tangential velocity from measurements in air at TUDA.}
    \label{fig:Recirc}
    \end{figure}

Figure\ \ref{fig:Recirc} shows tangential velocity distributions on the airfoil surface over dimensionless time.
 Velocities are extracted along three lines parallel to the surface and running tangential to the plate from the leading edge at $x/c=$ 0 to the trailing edge at $x/c=$1 for time steps of $\Delta (t/T)\approx$ 0.002. Fig.\ \ref{fig:Recirc} shows averaged values of these three lines that are located $y/c=$ 0.011 apart from each other with $y/c=$ 0.011 spacing to the surface. 
 
 Red colour coded areas indicate downstream velocities while blue colour coded areas velocity represents upstream fluid motion induced by the clockwise rotating LEV on the airfoil surface. So a LEV driven upstream velocity region can be clearly identified in Fig.\ \ref{fig:Recirc} convecting downstream on the airfoil. Upstream velocities on the airfoil surface occur from about $t/T\approx$ 0.1 for both evaluations. After emergence of the LEV a change of sign in velocity can be observed downstream of the vortex, which indicates the rear stagnation point of the flow at the rear of the LEV. For clarity it is marked with a green dashed line in Fig.\ \ref{fig:Recirc}. During the growth of the LEV, the stagnation point travels downstream until it reaches the trailing edge at $x/c=$ 1 (instant marked with a black square). From this instant on, recirculation of fluid from beneath the airfoil around the trailing edge is initiated. This instant agrees well between the two facilities: $t/T=$ 0.39 for BUAA and $t/T=$ 0.38 for TUDA. Reconsidering the instant at which circulation accumulation stops, identified from the LEV circulation evolution (at $t/T\approx$ 0.39), the recirculation of fluid occurs slightly before peak circulation. This shows that for the investigated baseline case circulation accumulation terminates due to fluid recirculation around the trailing edge for both evaluations, in accordance with \cite{Rival.2014}. Nevertheless, it should be noted that distinct secondary structures upstream of the LEV can be observed at later instants for $t/T\ge$ 0.2, indicated by additional changes of tangential velocity sign in horizontal direction in Fig.\ \ref{fig:Recirc}.

\section{Discussion}
The occurrence of secondary structures and their effect on the detachment process of the LEV is still not fully understood, as described in section~\ref{sec:sec_Intro}; their emergence and effect on the instant when the LEV stops accumulating circulation will be investigated in detail in this section. Based on the observed comparability of results obtained at the two facilities, each evaluation will consider results from both facilities to extend the covered parameter range.

\begin{figure}[!ht]
    \begin{subfigure}[t]{0.266\textwidth}
        \centering
        \includegraphics[width=\textwidth, trim = 40 7 85 21, clip]{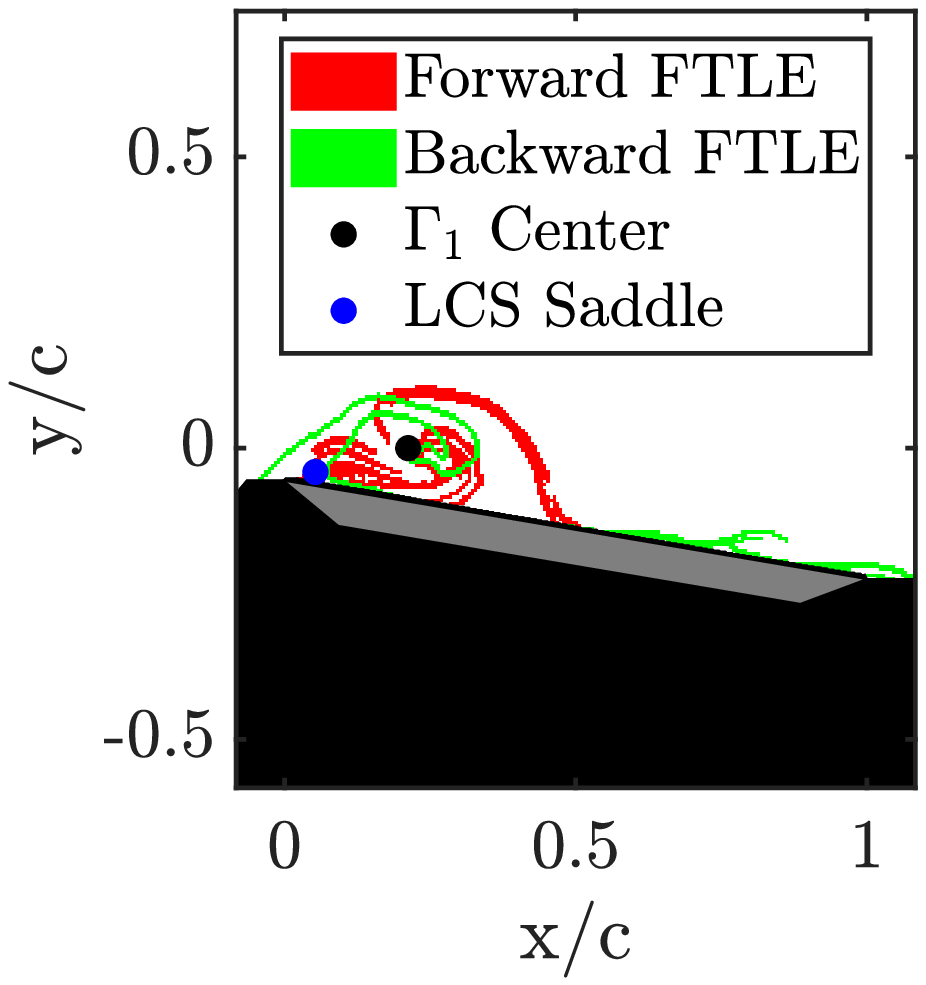}
        \caption{$t/T$ = 0.15}
        \label{fig:FTLERidge_tT0x15}
        \end{subfigure}
        \begin{subfigure}[t]{0.21\textwidth}
        \centering
        \includegraphics[width=\textwidth, trim = 95 7 85 21, clip]{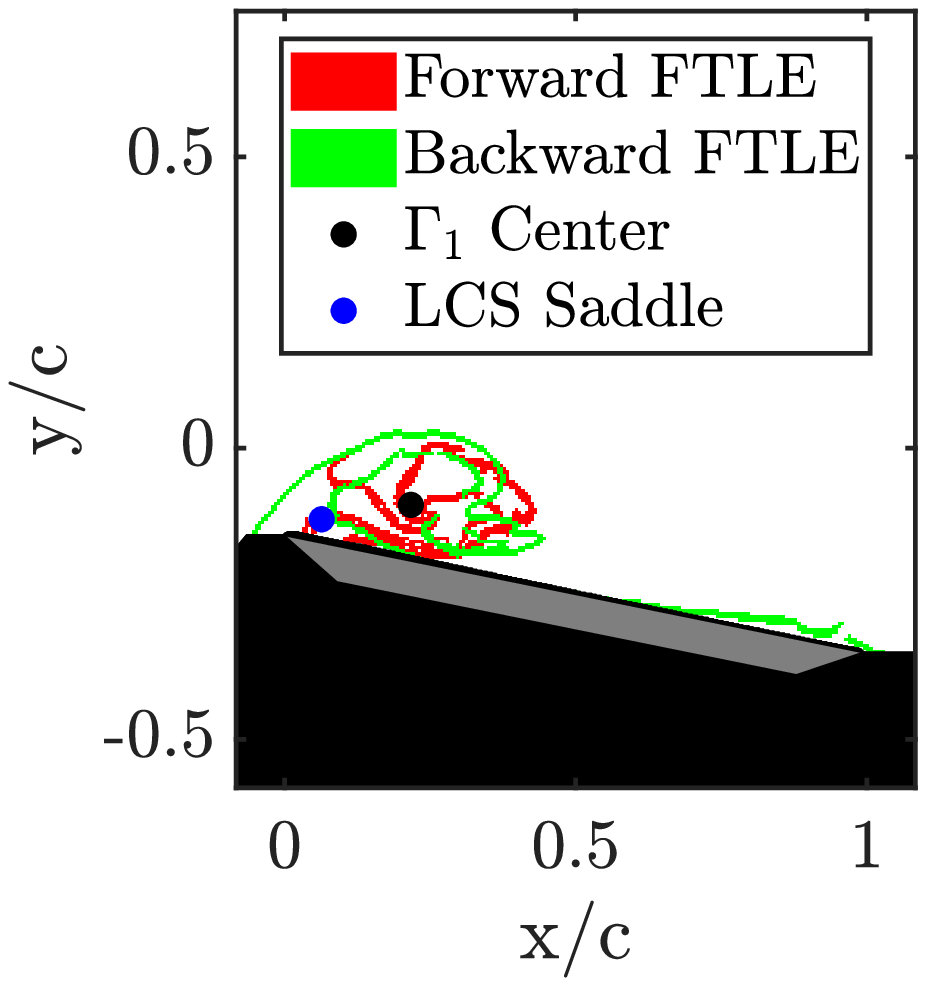}
        \caption{$t/T$ = 0.175}
        \label{fig:FTLERidge_tT0x2}
        \end{subfigure}
        \hfill
        \begin{subfigure}[t]{0.266\textwidth}
        \centering
        \includegraphics[width=\textwidth, trim = 40 7 85 21, clip]{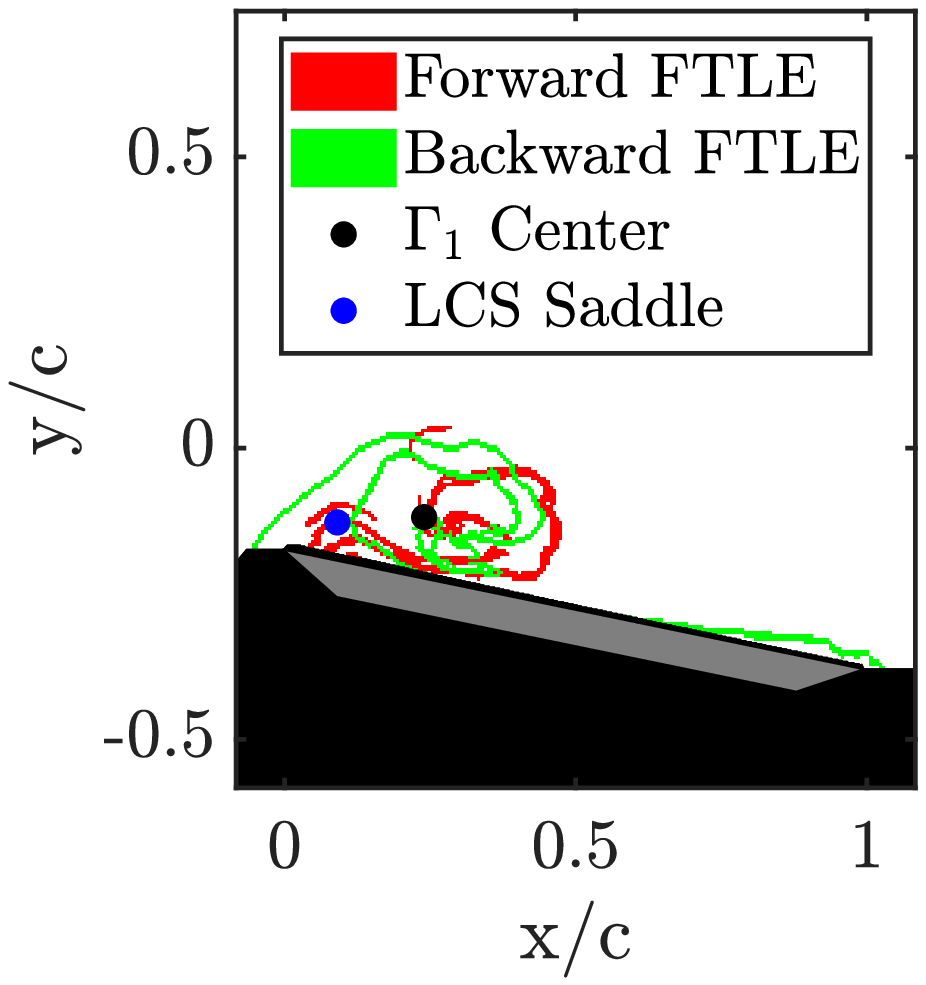}
        \caption{$t/T$ = 0.2}
        \label{fig:FTLERidge_tT0x21}
        \end{subfigure}
        \begin{subfigure}[t]{0.21\textwidth}
        \centering
        \includegraphics[width=\textwidth, trim = 95 7 85 21, clip]{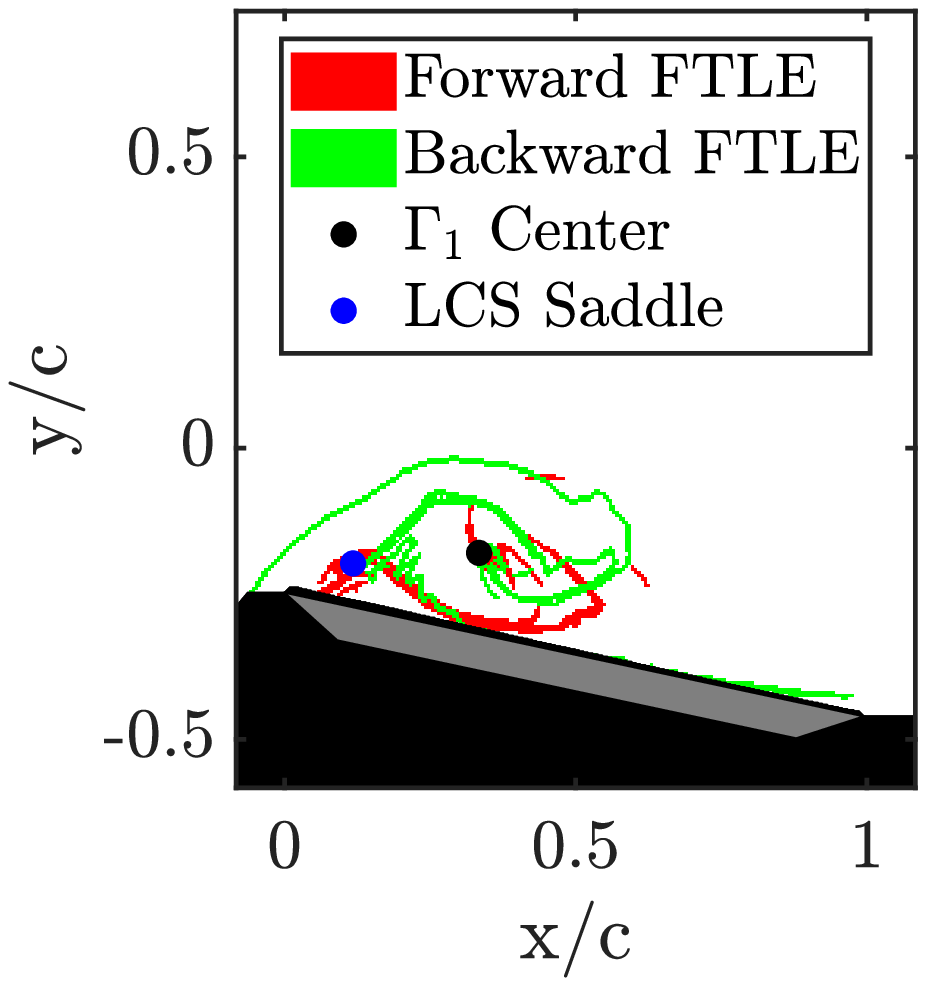}
        \caption{$t/T$ = 0.25}
        \label{fig:FTLERidge_tT0x25}
        \end{subfigure}
    \caption{Flow field in terms of repelling and attracting ridges for different dimensionless time instants obtained from forward and backward FTLE computations for the common baseline case (ID 3 from TUDA). FTLE ridges are obtained using a 80\,\% threshold of the respective maximum FTLE value in each frame. The airfoil is masked out in grey and the laser light shadow in black. The vortex center from $\Gamma_{1}$ criterion and the LCS saddle are marked as well.}
    \label{fig:FTLERidges}
\end{figure}

A combination of Lagrangian flow field analysis based on FTLE and Eulerian vortex characteristics is used for a detailed investigation of secondary structures and their effect on LEV characteristics, as discussed in section~\ref{SubSec:DataProc}. Fig.\ \ref{fig:FTLERidges} depicts FTLE flow fields in terms of attracting and repelling (backward and forward) FTLE ridges for different dimensionless time instants computed from velocity data obtained at the TUDA facility.
 Early in the downstroke at $t/T=$ 0.15, shown in Fig.\ \ref{fig:FTLERidge_tT0x15}, the LEV is confined at the top and to the rear by a shell-shaped repelling ridge (highlighted in red). At $t/T=$ 0.175 and for further time instants the confining ridge disappears, since its strength decreases. The leading edge shear layer feeding the LEV is demarcated in upstream direction by an attracting ridge (highlighted in green). In the downstream direction it is demarcated by an attracting ridge arising from the airfoil surface, forming a channel with the upstream demarcation line that curls up into the vortex. For $t/T=$ 0.15 and 0.175 there are no distinct flow structures visible at the leading edge, immediately above the airfoil. However from $t/T=$ 0.175 to 0.2 a region of fluid right above the airfoil surface at the leading edge emerges, which is isolated from the flow field by a repelling ridge. Fluid particles close to this ridge will depart from each other at future instances by becoming entrained into different topological structures, namely the main LEV and the secondary vortices, which are located close to the leading edge. The qualitative evolution of FTLE fields obtained from BUAA data was found to be in very good agreement regarding the observed topology.
 
\subsection{Secondary Structure Occurrence (SSO)}
\begin{figure}[ht]
    	\centering
       	\includegraphics[width=0.49\textwidth, trim = 5 5 35 12, clip]{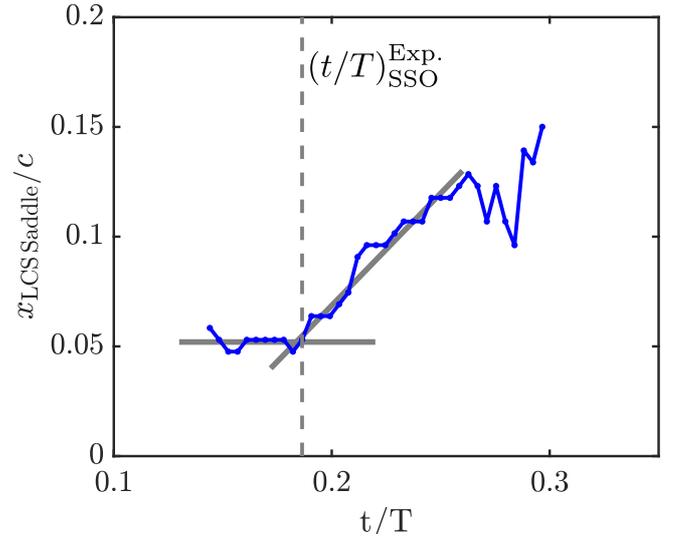}
       	\caption{Normalized streamwise LCS saddle position $x_{\mathrm{LCS\,Saddle}}/c$ as a function of dimensionless time $t/T$. The position is extracted in a plate-fixed frame of reference. The convection of the LCS saddle at different phases is approximated linear and indicated by solid grey lines. The derived instant of secondary structure occurrence $(t/T)_{\mathrm{SSO}}^{\mathrm{Exp.}}$ is marked with a vertical dashed line.}
       	\label{fig:LCS_XC}
    \end{figure}

In continuation of earlier efforts by \cite{Kissing.2020}, the occurrence of secondary structures is quantified with the approach introduced by \cite{Huang.2015}. This method identifies Lagrangian Coherent Structure (LCS) saddle points in the flow field by intersections of repelling and attracting ridges. The flow saddle point is indicated in Fig.\ \ref{fig:FTLERidges} as a blue point. As soon as secondary structures arise and grow, the LCS saddle point moves downstream as a result of the increased area covered by the structures. Therefore, the streamwise LCS saddle location in a plate-fixed frame of reference is extracted to evaluate secondary structure growth as depicted for ID 3 case from TUDA in Fig.\ \ref{fig:LCS_XC}.
After an initial phase where the LCS saddle remains close to the leading edge at about $x/c=$ 0.05, it starts to convect downstream shortly before $t/T=$ 0.2, as shown in Fig.\ \ref{fig:LCS_XC}, indicating an increase of the area covered by secondary structures and thus their growth. This evolution is in agreement with the observed secondary structure onset from FTLE fields in Fig.\ \ref{fig:FTLERidges}. To determine the instant of convection increase and thus secondary structure occurrence, the LCS saddle location is approximated first order separately from $t/T=$ 0.15 to 0.18 and from 0.18 to 0.25. The dimensionless time instant of secondary structure occurrence (SSO), denoted as $(t/T)_{\mathrm{SSO}}^{\mathrm{Exp.}}$, can be determined by the intersection of both convection slopes. It is marked in Fig.\ \ref{fig:LCS_XC} by a vertical dashed line. The convection of the LCS saddle evaluated from FTLE fields obtained from BUAA for baseline case data was found to occur about 3\,\% earlier than from TUDA with respect to the downstroke period, which is considered to be very good agreement. Subsequently, $(t/T)_{\mathrm{SSO}}^{\mathrm{Exp.}}$ was evaluated for all investigated cases. 

\begin{figure}[ht]
    	\centering
       	\includegraphics[width=0.49\textwidth, trim = 0 0 0 0, clip]{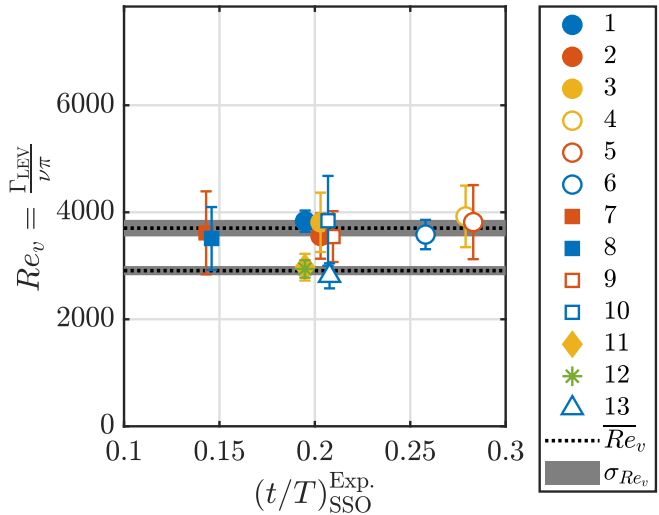}
       	\caption{Vortex Reynolds number $Re_{v}$ at secondary structure onset $(t/T)_{\mathrm{SSO}}^{\mathrm{Exp.}}$. Measurements at TUDA (ID 1 to 10) and measurements at BUAA (ID 11 to 13) are used to obtain separate mean vortex Reynolds numbers $\overline{Re}_{v}$ and their respective standard deviations $\sigma_{Re_{v}}$, indicated as dashed lines and grey patches; see Table\ \ref{tab:AllCasesTUDA} and \ref{tab:AllCasesBUAA} for the respective parameter combinations of ID 1-13.}
       	\label{fig:Re_Vort}
    \end{figure}

The circulation of the LEV from vortex identification evaluations was additionally used to compute the vortex Reynolds number $Re_{v} = \Gamma_{\mathrm{LEV}} / \nu \pi$ proposed by \cite{Doligalski.1994} at the determined secondary structure onset. The vortex Reynolds number represents the tendency of the boundary layer below the vortex on the airfoil surface to respond to the vortex induced pressure gradient in a viscous fashion, leading to an eruption of fluid and subsequent formation of secondary structures. For low $Re_{v}$ this viscous response is suppressed while it is triggered at higher values. $Re_{v}$ is shown as a function of $(t/T)_{\mathrm{SSO}}^{\mathrm{Exp.}}$ in Fig.\ \ref{fig:Re_Vort}.
Considering the secondary structure onset for cases with $\hat{\alpha}_{\mathrm{eff}}$ = 30$\, ^{\circ}$ from TUDA, represented by the filled markers (ID 1; 2; 3; 7; 8), it can be observed that secondary structures arise at earlier dimensionless times for lower $k$ (square markers) than for a higher ones (circle markers). The Strouhal number has a minor influence on the onset instant (different colours). Similar trends can be observed for $\hat{\alpha}_{\mathrm{eff}}$ = 20$\, ^{\circ}$ cases, indicated by the open markers (ID 4; 5; 6; 9; 10), with a general temporal lag. The instants of secondary structure onset from BUAA agree well with the TUDA results, considering cases of the same parameters (ID 3 and 11 as well as ID 10 and 13). The observation of a minor influence of the Strouhal number is additionally confirmed by the ID 12 case agreement with ID 3 and 11 cases, since it denotes a case with parameters identical to them but with a higher $St$ of 0.16.

For parameters investigated in this study, secondary structures occur when the vortex Reynolds number reaches a threshold within a very narrow band for each setup, although the mean vortex Reynolds numbers $\overline{Re}_{v}$ differ between both. For TUDA results, secondary structures occur for $Re_{v}$ between 3,500 and 3,900 with $\overline{Re}_{v}=$ 3,700 and a standard deviation $\sigma_{Re_{v}}$ of about 150. At BUAA they arise for $\overline{Re}_{v}=$ 2900 with a smaller $\sigma_{Re_{v}}$ of about 90. Reconsidering the comparison of circulation evolution for the common baseline case from Fig.\ \ref{fig:CircBase_Gamma2}, where an offset for earlier instants could be attributed to the inclusion of the leading edge shear layer for the TUDA evaluations, the higher $Re_{v}$ from TUDA results can be expected and attributed to the vortex identification methodology.

Based on a known threshold of the onset of secondary structures in terms of $\overline{Re}_{v}$, which is in turn dependent on the LEV circulation, their onset could be predicted if the accumulation of circulation of the LEV could be modelled. This prediction would be valuable since future attempts to delay the LEV detachment will focus on the suppression of these secondary structures, which requires their onset to be known to specifically delay or suppress their occurrence. \cite{Wong.2015} derive an expression for the rate of circulation accumulation of the LEV,  $\dot{\Gamma}_{\mathrm{LEV}}$, in which it is proportional to the square of the effective inflow velocity on the airfoil $u_{\mathrm{eff}}$. This scaling was found to capture $\Gamma_{\mathrm{LEV}}$, experimentally obtained via vortex identification, precisely for cases where the LEV forms very early in the downstroke.

\begin{figure}[ht]
    	\centering
       	\includegraphics[width=0.49\textwidth, trim = 0 0 0 0, clip]{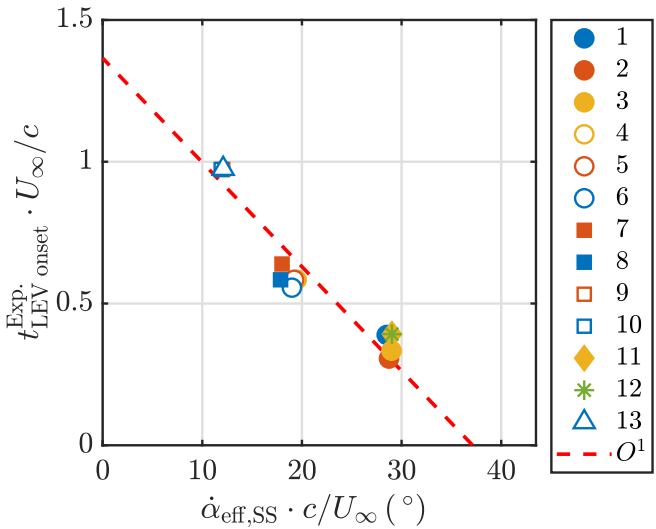}
       	\caption{LEV onset delay $t_{\mathrm{LEV\,onset}}^{\mathrm{Exp.}}$ as a function of the rate of change of the effective inflow angle on the airfoil during static stall angle $\dot{\alpha}_{\mathrm{eff,SS}}$. A linear approximation of the scaling is indicated as dashed red line; see Table\ \ref{tab:AllCasesTUDA} and \ref{tab:AllCasesBUAA} for the respective parameter combinations of ID 1-13.}
       	\label{fig:AlphaDotStaticStall_vs_tonset}
    \end{figure}

During the comparison of the circulation evolution predicted by the aforementioned model and measured evolution, significant deviations were encountered for several cases. In these cases the LEV was found to emerge delayed in respect to the motion start, caused by a delayed roll-up of the leading edge shear layer. This led to a temporal offset of the measured LEV circulation with respect to the predicted value, although the slope of circulation accumulation was still in very good agreement. To account for the observed delayed LEV formation, \cite{Mulleners.2012} developed a scaling for the temporal LEV formation lag with respect to the motion start $t_{\mathrm{LEV\,onset}}^{\mathrm{Exp.}}$, which is tested as a potential solution to adapt the aforementioned circulation flux model. It is based on the assumption that the rate of change of the inflow angle on the airfoil during the period when the inflow angle exceeds the static stall angle is responsible for the temporal lag of LEV formation. Fig.\ \ref{fig:AlphaDotStaticStall_vs_tonset} shows the rate of change of the effective inflow angle during this period $\dot{\alpha}_{\mathrm{eff,SS}}$ as a function of the LEV onset delay $t_{\mathrm{LEV\,onset}}^{\mathrm{Exp.}}$, determined by visual inspection of FTLE fields.
By comparing the observed LEV emergence from TUDA and BUAA evaluations with a linear approximation, indicated by the dashed red line, good agreement in terms of a linear scaling by $\dot{\alpha}_{\mathrm{eff,SS}}$ can be observed.

\begin{figure}[ht]
    	\centering
       	\includegraphics[width=0.49\textwidth, trim = 0 0 0 0, clip]{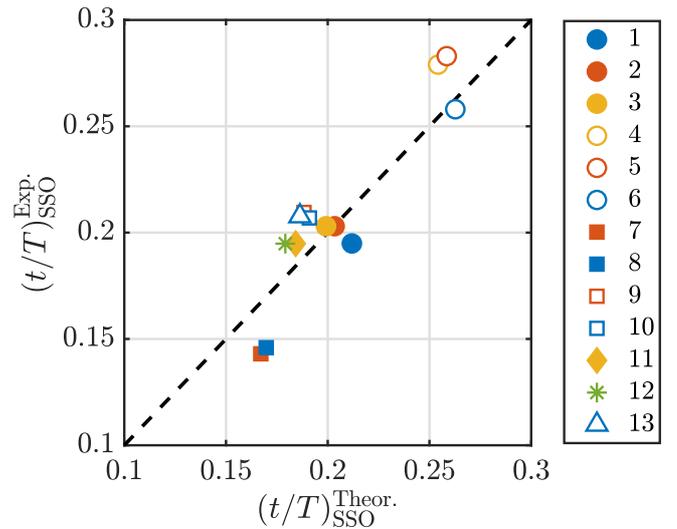}
       	\caption{Modelling secondary structure occurrence $(t/T)_{\mathrm{SSO}}^{\mathrm{Theor.}}$ based on a vortex Reynolds number threshold, taking the delay of LEV formation in circulation approximation into account; see Table\ \ref{tab:AllCasesTUDA} and \ref{tab:AllCasesBUAA} for the respective parameter combinations of ID 1-13.}
       	\label{fig:SecStructOnset_ExpTheo}
    \end{figure}

Accordingly, the circulation flux model by \cite{Wong.2015} is adapted to account for the delayed vortex formation. This is done by offsetting temporal information used to integrate $\dot{\Gamma}_{\mathrm{LEV}}$ from the model according to the linearly approximated $t_{\mathrm{LEV\,onset}}^{\mathrm{Exp.}}$. By using separate circulation threshold values, derived from $\overline{Re}_{v}$ thresholds for TUDA and BUAA, the dimensionless time instant of expected secondary structure onset $(t/T)_{\mathrm{SSO}}^{\mathrm{Theor.}}$ is obtained from the adapted circulation flux model. It is compared to the experimentally observed secondary structure occurrence $(t/T)_{\mathrm{SSO}}^{\mathrm{Exp.}}$ in Fig.\ \ref{fig:SecStructOnset_ExpTheo}.
The maximum deviation between the predicted and measured secondary structure occurrence $\Delta(t/T)$ is 0.0248 (equivalent to 4.9\% of the downstroke period), which is considered as a reasonably accurate prediction.

\subsection{Secondary Structure Effect on LEV Detachment}
In light of the discrepancies regarding the LEV detachment mechanism encountered in literature (cf. section\ \ref{sec:sec_Intro}), the role of secondary structures in the detachment process is discussed below. This discussion will focus on secondary structure effects on the termination of circulation accumulation of the LEV to clarify their role as a trigger of secondary topological structures. It should be noted that the LEV detachment can also be defined in terms of the lift force decrease, which however, has not been measured in this study.

\begin{figure}[ht]
    	\centering
       	\includegraphics[width=0.49\textwidth, trim = 0 0 0 0, clip]{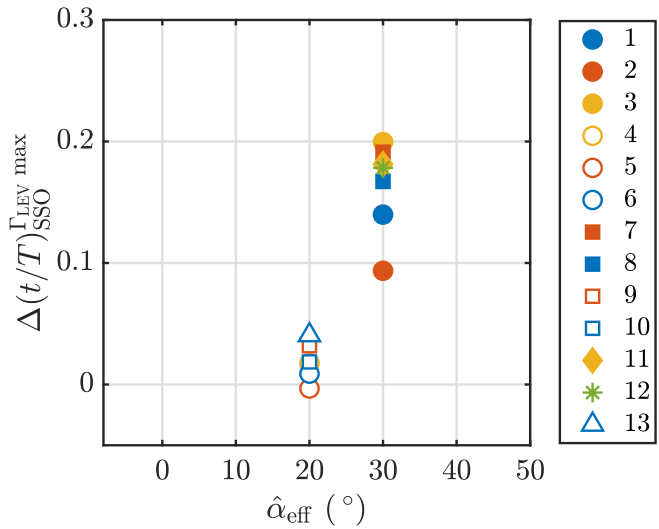}
       	\caption{Dimensionless time lag between the onset of secondary structures and the instant of maximum LEV circulation $\Delta (t/T)_{\mathrm{SSO}}^{\mathrm{\Gamma_{LEV}\,max}}$ to evaluate the concurrency of both events, shown over of the effective inflow angle amplitude on the airfoil $\hat{\alpha}_{\mathrm{eff}}$; see Table\ \ref{tab:AllCasesTUDA} and \ref{tab:AllCasesBUAA} for the respective parameter combinations of ID 1-13.}
       	\label{fig:tTSecStruct_vs_CircAccStop}
    \end{figure}

To identify parameter sets where circulation accumulation of the LEV stops as a direct consequence of secondary structure emergence, the concurrency of both events is compared. This is done with the aid of the dimensionless time lag between secondary structure onset ($(t/T)_{\mathrm{SSO}}^{\mathrm{Exp.}}$) and maximum LEV circulation ($(t/T)^{\mathrm{\Gamma_{LEV}\,max}}$), denoted as $\Delta (t/T)_{\mathrm{SSO}}^{\mathrm{\Gamma_{LEV}\,max}}$. It is shown in Fig.\ \ref{fig:tTSecStruct_vs_CircAccStop} as a function of the effective inflow angle amplitude on the airfoil $\hat{\alpha}_{\mathrm{eff}}$.
The instant of secondary structure onset is determined by topological FTLE field evaluations according to the methodology introduced along the discussion of Fig.\ \ref{fig:LCS_XC} and the instant of LEV peak circulation from vortex identification results, indicated by the dashed lines in Fig.\ \ref{fig:CircBase_Gamma2}. For parameter sets with $\hat{\alpha}_{\mathrm{eff}}=$ 20$\, ^{\circ}$, highlighted by open symbols, a small time lag $\Delta (t/T)_{\mathrm{SSO}}^{\mathrm{\Gamma_{LEV}\,max}}$ between 0.003 and 0.04 can be observed. On the other hand the time lag is generally larger for cases with $\hat{\alpha}_{\mathrm{eff}}$ = 30$\, ^{\circ}$, represented by the filled markers and the star marker. For all cases with $\hat{\alpha}_{\mathrm{eff}}$ = 20$\, ^{\circ}$, fluid recirculation around the trailing edge occurs significantly after the LEV reaches its peak circulation, which in turn renders fluid recirculation an insuficient indicator of the end of the accumulation of circulation of the LEV. These observations suggest that for lower effective angle of attack amplitudes the emergence of secondary structures causes the LEV to stop circulation accumulation, independent of fluid recirculation around the trailing edge.

A potential scenario explaining how secondary structures cause the LEV to stop accumulating circulation is an interaction of the shear layer fluid from the leading edge of the airfoil with secondary structures, which takes place before the shear layer fluid reaches the LEV. An interaction of the negatively signed vortical fluid from the shear layer with the positively signed of secondary structures would presumably result in cross-annihilation of vorticity, as described by \cite{Wojcik.2014}. Additionally, the shear layer of high velocity could push secondary structures and the main LEV downstream if a low shear layer angle directs fluid directly into those structures. 

The geometric factors determining whether an interaction of any kind is enabled are the evolution of position and size of secondary structures in combination with the shear layer angle. As soon as secondary structures arise, they will grow and consequently cover a larger area. If the shear layer angle does not increase further from the instant of secondary structure onset, an interaction of both due to growing secondary structures is a direct consequence. For lower peak inflow angles on the airfoil the shear layer angle is assumed to reach lower peak values and an interaction of secondary structures with the shear layer occurs earlier. This could explain the observed concurrency of secondary structure onset and the termination of circulation accumulation of the LEV for cases where $\hat{\alpha}_{\mathrm{eff}}$ is 20$\, ^{\circ}$ as shown in Fig.\ \ref{fig:tTSecStruct_vs_CircAccStop}.

\begin{figure}[ht]
    	\centering
       	\includegraphics[width=0.49\textwidth, trim = 0 0 0 0, clip]{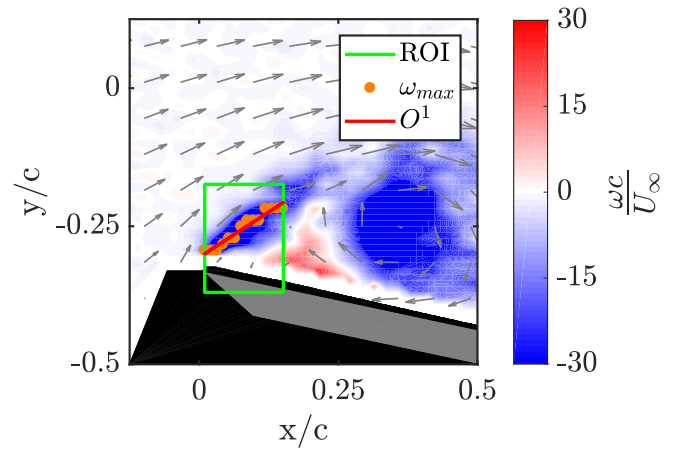}
       	\caption{Extraction methodology of the shear layer angle after separation from the leading edge $\alpha_{\mathrm{SL}}$. The normalized vorticity field around the leading edge region is depicted at $t/T=$ 0.26 from ID 8 case. Highest vorticity values ($\omega_{\mathrm{max}}$) within the region of interest (ROI) are identified to obtain a first order approximated line ($O^{1}$). The inflow is from the left, the airfoil is masked out in grey and the laser light shadow in black.}
       	\label{fig:SLEffExtraction}
\end{figure}

\begin{figure}[ht]
    	\centering
       	\includegraphics[width=0.49\textwidth, trim = 0 0 0 0, clip]{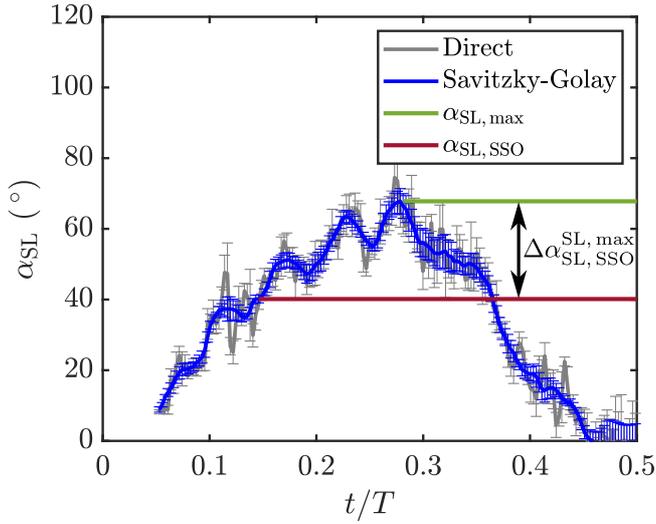}
       	\caption{Ensemble averaged evolution of the extracted shear layer angle after separation from the leading edge $\alpha_{\mathrm{SL}}$ over dimensionless time $t/T$ for the ID 8 case from TUDA. Error bars represent the standard deviation of single run ensemble averaging. The shear layer angle during secondary structure onset $\alpha_{\mathrm{SL,\,SSO}}$ and the overall maximum angle $\alpha_{\mathrm{SL,\,max}}$ in addition to their difference $\Delta \alpha_{\mathrm{SL,\,SSO}}^{\mathrm{SL,\,max}}$ are also indicated.}
       	\label{fig:SLEffEvolution}
    \end{figure}

To test this hypothesis, the influence of the shear layer angle after separation from the leading edge $\alpha_{\mathrm{SL}}$ will be investigated as a potential factor causing the simultaneous cessation of circulation accumulation and onset of secondary structures for $\hat{\alpha}_{\mathrm{eff}}$ = 20$\, ^{\circ}$. The methodology used to extract $\alpha_{\mathrm{SL}}$ from vorticity fields is based on the fact that the leading edge shear layer is characterized by very high vorticity values. This methodology is illustrated exemplary for the ID 8 Case at $t/T =$ 0.26 in Fig.\ \ref{fig:SLEffExtraction}.
Maximum vorticity values, denoted as $\omega_{\mathrm{max}}$ and indicated as orange dots in Fig.\ \ref{fig:SLEffExtraction}, are extracted in a square, plate-fixed region of interest (ROI) that spans from the leading edge downstream up to the region of secondary structures. Final $\alpha_{\mathrm{SL}}$ evolution is obtained from the angle between a first order fitted line ($O^1$) of the extracted maximum value locations, indicated by a red line in Fig.\ \ref{fig:SLEffExtraction}, and the airfoil surface. To exclude large fluctuations of the shear layer angle, its evolution is smoothed using a second order Savitzky-Golay filter of 15 frames width (corresponding to $\Delta (t/T)=$ 0.1). After extraction of the angle from single runs, it is ensemble averaged for each parameter set. An exemplary evolution of $\alpha_{\mathrm{SL}}$ for the ID 8 case from TUDA is depicted in Fig.\ \ref{fig:SLEffEvolution}, where the direct ensemble averaged angle evolution is shown in grey and the smoothed in blue with error bars, indicating the standard deviation between single runs.
    
\begin{figure}[ht]
    	\centering
       	\includegraphics[width=0.49\textwidth, trim = 0 0 0 0, clip]{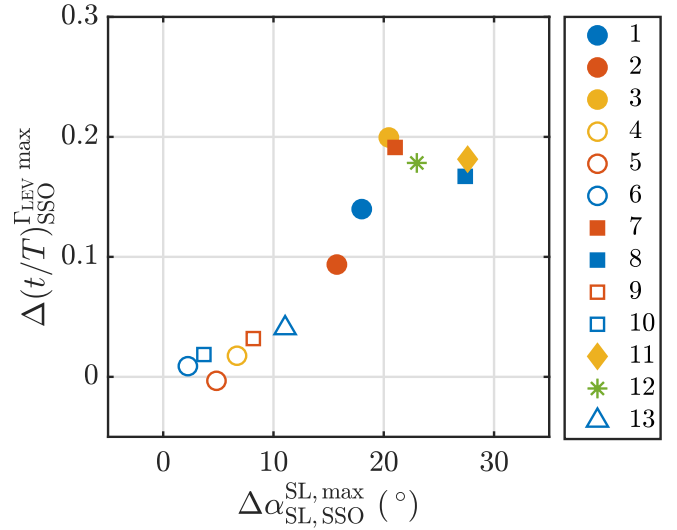}
       	\caption{Correlation of the temporal concurrency between circulation accumulation stop of the LEV and secondary structure occurrence $\Delta (t/T)_{\mathrm{SSO}}^{\mathrm{\Gamma_{LEV}\,max}}$ with the shear layer angle difference at secondary structure onset and the maximum angle $\Delta \alpha_{\mathrm{SL,\,SSO}}^{\mathrm{SL,\,max}}$; see Table\ \ref{tab:AllCasesTUDA} and \ref{tab:AllCasesBUAA} for the respective parameter combinations of ID 1-13.}
       	\label{fig:DeltaAlpha_tTSecStructMax}
    \end{figure}

The geometric coefficient that determines whether secondary structures can grow without interacting with the shear layer fluid after their emergence is the difference between the shear layer angle at secondary structure onset $\alpha_{\mathrm{SL,\,SSO}}$ and the maximum shear layer angle $\alpha_{\mathrm{SL,\,max}}$. This difference is denoted as $\Delta \alpha_{\mathrm{SL,\,SSO}}^{\mathrm{SL,\,max}}$ and highlighted in Fig.\ \ref{fig:SLEffEvolution} by a double-sided arrow. For larger values of $\Delta \alpha_{\mathrm{SL,\,SSO}}^{\mathrm{SL,\,max}}$ the shear layer angle increases more after secondary structure onset and an interaction is unlikely to take place immediately, while it is likely to occur directly if the angle does not increase further. Fig.\ \ref{fig:DeltaAlpha_tTSecStructMax} shows the concurrency of the emergence of secondary structures and the peak circulation of the LEV as a function of $\Delta \alpha_{\mathrm{SL,\,SSO}}^{\mathrm{SL,\,max}}$.
In cases with $\hat{\alpha}_{\mathrm{eff}}$ = 20$\, ^{\circ}$, indicated by open symbols, $\Delta \alpha_{\mathrm{SL,\,SSO}}^{\mathrm{SL,\,max}}$ is generally lower, so the shear layer angle at the leading edge increases less from the instant of secondary structure onset before reaching its maximum. This trend occurs concurrent with lower $\Delta (t/T)_{\mathrm{SSO}}^{\mathrm{\Gamma_{LEV}\,max}}$ values, indicating a circulation accumulation stop of the vortex as a more direct consequence of secondary structure emergence. In contrast, the shear layer angle increases significantly  after secondary structure emergence before reaching its maximum for $\hat{\alpha}_{\mathrm{eff}}$ = 30$\, ^{\circ}$ (filled markers). The temporal lag of circulation accumulation stop of the LEV after secondary structure emergence $\Delta (t/T)_{\mathrm{SSO}}^{\mathrm{\Gamma_{LEV}\,max}}$ correlates with $\Delta \alpha_{\mathrm{SL,\,SSO}}^{\mathrm{SL,\,max}}$ for either case.

Regarding the hypothesis of an interaction of secondary structures and the leading edge shear layer, this shows that if the shear layer angle increases further after secondary structure occurrence, an interaction is delayed and the vortex accumulates circulation for a longer period. This highlights the effective shear layer angle as an important factor that influences when the LEV ceases to accumulate circulation.

\section{Conclusions}
In this study an approach to establish comparability of flow fields and LEV characteristics on a pitching and plunging flat plate obtained from facilities working with air and water as media is developed and validated. An order of magnitude difference in viscosity of the media extends the attainable parameter space in terms of dimensionless parameters.

The definition of a common baseline case was found to require careful consideration of the viable parameter range due to inherently different time scales of the airfoil motion caused by the difference in viscosity. Flow fields and topological evolution agreed well in air and water, but the LEV circulation differed. An intermittent inclusion of the leading edge shear layer into the LEV domain was identified as the source for deviations of the computed circulation. This highlights that small deviations in velocity information can affect circulation evaluations via the identified vortex domain used for circulation computation, even when using the same correlation algorithm. As already stated in section \ref{sec:sec_FacMeth}, data sets of baseline cases from TUDA and BUAA are also available online as reference cases (http://dx.doi.org/10.25534/tudatalib-168).

An investigation of secondary structures ahead of the LEV with the aid of a Lagrangian flow field analysis by FTLE ridges allowed a precise identification of their onset and effect on the LEV detachment process. The combination of both facilities enabled investigations of secondary structures over an extended parameter range, including variations of the reduced frequency, the Strouhal number and the effective angle of attack amplitude.

Secondary structures were found to emerge at similar vortex Reynolds numbers, computed from LEV circulation, at each setup and for all investigated parameters. Secondary structure occurrence, which is triggered by a viscous response of the boundary layer below the airfoil, is thus governed by the vortex circulation.

With the aid of existing circulation flux models from literature and their adaptation to account for a delayed LEV formation, a model to predict the temporal occurrence of secondary structures was developed and found to be in agreement with the observed onset.

For cases with a lower effective angle of attack amplitude, the LEV was found to stop accumulating circulation in close temporal correlation with secondary structure emergence. The temporal concurrence of secondary structure onset and the circulation accumulation stop was found to closely correlate with the leading edge shear layer angle increase after secondary structure onset.

Termination of circulation accumulation of the LEV due to the occurrence of secondary structures implies that a suppression of their onset or growth is a promising approach for future flow control strategies targeting a prolonged LEV growth phase. The timing of secondary structure suppression can be optimized using the developed model for secondary structure onset.



\begin{acknowledgements}
The authors wish to acknowledge financial support of  the Sino-German Center and the Deutsche Forschungsgemeinschaft through the project TR 194/55-1: ``Flow Control for Unsteady Aerodynamics of Pitching/Plunging Airfoils". The authors would also like to extend their appreciation to the workshop staff in Darmstadt for their assistance and professional support  for these experiments.
\end{acknowledgements}
\bibliographystyle{spbasic}      
\bibliography{Insights_into_Leading-Edge_Vortex_Formation_and_Detachment}   

\end{document}